\documentclass[a4paper,11pt]{article}
\pdfoutput=1 

\usepackage{jcappub} 
\usepackage[toc,page]{appendix}
\usepackage{subcaption}
\usepackage{amsmath}
\usepackage{graphicx}                
\usepackage{multirow}
\usepackage[T1]{fontenc} 
\usepackage[table]{xcolor}
\usepackage{float}

\usepackage{orcidlink}
\usepackage{booktabs}
\usepackage{array}

\hypersetup{pdftitle={UV artefacts in ultra-slow-roll models of inflation}
}
\usepackage{bm}
\renewcommand{\arraystretch}{1.15}

\title{\boldmath UV artefacts in ultra-slow-roll models of inflation}

\author[a]{Gerald Barnert,\,$^{\text{\orcidlink{0000-0001-5411-014X}}}$}
\author[b,a]{Laura Iacconi,\,$^{\text{\orcidlink{0000-0002-1152-3056}}}$}
\author[a]{Hooshyar Assadullahi,\,$^{\text{\orcidlink{0000-0003-0837-0199}}}$}
\author[a, c, d]{Kazuya Koyama\,$^{\text{\orcidlink{0000-0001-6727-6915}}}$}
\author[a]{and David Wands\,$^{\text{\orcidlink{0000-0001-9509-8386}}}$}

\affiliation[a]{Institute of Cosmology \& Gravitation, University of Portsmouth,\\ Burnaby Road, Portsmouth, PO1 3FX, U.K.}
\affiliation[b]{Astronomy Unit, Queen Mary University of London, \\ Mile End Road, London, E1 4NS, U.K.}
\affiliation[c]{Kavli IPMU (WPI), UTIAS, The University of Tokyo, \\ Kashiwa, Chiba 277-8583, Japan}
\affiliation[d]{Yukawa Institute for Theoretical Physics, Kyoto University, \\ Kyoto 606-8502, Japan}

\emailAdd{gerald.barnert@port.ac.uk}
\emailAdd{l.iacconi@qmul.ac.uk}
\emailAdd{hooshyar.assadullahi@port.ac.uk}
\emailAdd{kazuya.koyama@port.ac.uk}
\emailAdd{david.wands@port.ac.uk}

\abstract{
Within single-field inflation, primordial black hole and scalar-induced gravitational wave production from enhanced primordial perturbations typically requires a transient non-attractor phase, such as ultra-slow roll. 
We investigate the physical consistency of modeling such scenarios through analytical Hubble-flow parametrisation. 
By reconstructing the underlying scalar field potential, we show that even slow transitions in the slow-roll parameters can hide sharp, localised spikes in higher-order derivatives of the potential at the transition from ultra-slow-roll to slow-roll. 
These are typically not found in analytic potentials. 
To evaluate the impact of these structures, we implement a UV-filtering procedure based on discrete Fourier transform to systematically suppress high frequency modes in field space in both classes of models. 
We find that the filter effectively removes sharp features in Hubble-flow-derived potentials.
As a consequence, we show that UV-filtered models typically respect Wands duality invariance as the field evolves back from ultra-slow roll to slow roll. 
Beyond linear perturbation theory, the introduction of spurious UV effects might affect other observables, such as non-Gaussianity and loop contributions. 
Our results thereby question the robustness of simple analytical Hubble-flow parametrisation for modeling inflationary models with a transient non-attractor phase.
}

\begin{document}
	\maketitle
	\flushbottom
	
\section{Introduction}
\label{sec: introduction}

Cosmological inflation was proposed as a solution to several issues within the standard hot Big Bang cosmology~\cite{Starobinsky:1979ty, Starobinsky:1980te, Guth:1980zm}.
The simplest models of inflation feature a canonical scalar field, the inflaton, minimally coupled to gravity and endowed with a potential characterised by a sufficiently flat region.
The ensuing slow-roll (SR) dynamics drives the accelerated expansion of the background spacetime~\cite{Linde:1981mu, Albrecht:1982wi}.
Moreover, quantum fluctuations of the inflaton provide a mechanism for explaining the origin of primordial perturbations observed in the Cosmic Microwave Background (CMB)~\cite{Planck:2018vyg, Planck:2018jri, AtacamaCosmologyTelescope:2025blo, SPT-3G:2025bzu}.

The inflationary background dynamics is described by the Hubble parameter $H \equiv \dot a/a$ and its time dependence. 
Here $a(t)$ is the scale factor and a dot indicates a derivative with respect to cosmic time, $t$. 
Departures from the de Sitter solution $a(t) \propto e^{Ht}$ are commonly characterised through a hierarchy of SR parameters 
\begin{equation}
    \epsilon_1 \equiv -\frac{\mathrm{d} \ln H}{\mathrm{d} N}, \quad \epsilon_{i+1} \equiv \frac{\mathrm{d} \ln \epsilon_i}{\mathrm{d}N} \; \forall \; i\geq1 \;,
\end{equation}
where $N$ is the number of e-folds, defined by $\mathrm{d}N \equiv H\,\mathrm{d}t$. 
SR dynamics is realised when all SR parameters are small, leading to an attractor solution in which the inflaton velocity is determined by the slope of the potential, $\dot \phi \simeq -V_{\phi}/(3H)$, where $V_{\phi}\equiv \mathrm{d}V(\phi)/\mathrm{d}\phi$. 

On large scales $(k \in [10^{-3},\, 1]\,\mathrm{Mpc}^{-1})$, CMB observations indicate that primordial perturbations are adiabatic, nearly Gaussian, and characterised by an almost scale-invariant power spectrum $\mathcal{P}_{\zeta}(k)$, with amplitude $\mathcal{P}_{\zeta} \sim 10^{-9}$~\cite{Planck:2018vyg, Planck:2018jri, AtacamaCosmologyTelescope:2025blo, SPT-3G:2025bzu}.
Here we have introduced the dimensionless power spectrum, obtained from the Fourier transform of the 2-point correlation function of the curvature perturbation, $\zeta$,
\begin{equation}
    \langle \hat\zeta_\mathbf{k} \hat\zeta_{\mathbf{k'}}\rangle 
    \equiv
    (2\pi)^3 \,
    \delta^{(3)}(\mathbf{k}+\mathbf{k'}) \,
    \frac{2\pi^2}{k^3}\mathcal{P}_{\zeta}(k) \;. 
\end{equation}
Standard single-field SR inflation successfully explains CMB observations.

On much shorter scales $(k \gg 1\,\mathrm{Mpc}^{-1})$, the statistics of primordial perturbations are less constrained, and could potentially deviate from what observed on large scales.  
For example, if $\zeta$ is strongly enhanced (e.g. $\mathcal{P}_{\zeta} \sim 10^{-3}$~\cite{Gow:2020bzo}) it might lead to the formation of primordial black holes (PBHs) during radiation domination~\cite{Hawking:1971ei,Carr:1974nx}, as well as generating observable scalar-induced gravitational waves~\cite{Domenech:2021ztg}.
PBHs have been investigated as dark-matter candidates, contributors to observed gravitational wave events and progenitors of super-massive black holes observed in the center of galaxies~\cite{Carr:2026hot}. 

Such dramatic enhancement of $\mathcal{P}_{\zeta}$ cannot be achieved within standard SR attractor models. 
Instead, a transient departure from attractor dynamics is required. 
This can be understood by considering the behaviour of $\zeta$ on super-horizon scales
\begin{equation}
    \label{eqn: zeta super-horizon solution1}
    \zeta(N)_{k\ll aH} = c_1 + c_2 \int \mathrm{d}N' e^{-\int \mathrm{d}N'' (3-\epsilon_1+\epsilon_2)} \;. 
\end{equation}
In standard SR inflation $(|\epsilon_i| \ll 1 \; \forall \,i \geq 1)$, the second solution decays exponentially, leaving only the constant adiabatic mode. 
However, for non-attractor dynamics $(\epsilon_1\ll1, \; \epsilon_2<-3)$ the would-be-decaying mode instead grows exponentially, leading to super-horizon evolution of $\zeta$. 
Such dynamics can be accommodated within single-field models of inflation as a transient departure from the SR attractor regime, long after the CMB modes exited the horizon~\cite{Leach:2001zf}.  
The transient non-attractor phase is then followed by an attractor era, which quenches the growth of perturbations and leads to the end of inflation. 

Non-attractor dynamics can be realised if the inflationary potential features a nearly-flat plateau, an inflection point, or other localised structures such as a bump or a dip (see, e.g., Ref.~\cite{Cole:2023wyx} for a sample of models).
Dynamics of this kind constitutes a sub-class of constant-roll inflation~\cite{Kinney:2005vj, Motohashi:2025qgd}, characterised by $\ddot \phi - \beta H \dot \phi\simeq 0$, with $\beta$ constant, and $\epsilon_2\approx 2 \beta$.  
A well studied example of non-attractor, constant-roll dynamics is ultra-slow-roll (USR)~\cite{Ivanov:1994pa, Yokoyama:1998pt, Inoue:2001zt}, which arises when $V_\phi\simeq 0$ and the inflaton evolves according to $\ddot \phi + 3H \dot \phi\simeq 0$ ($\beta=-3$).
In this case, $\epsilon_1 \propto a^{-6}$, leading to $\epsilon_2 \approx -6$. 

In this paper, we study a widely used parametrisation~\cite{Taoso:2021uvl, Franciolini:2022pav, Franciolini:2023agm, Caravano:2024moy, Caravano:2025diq, Caravano:2026hca} that allows one to interpolate between attractor and non-attractor dynamics by specifying the time-dependence of the SR parameters, rather than deriving the background and perturbation evolution from a specific $V(\phi)$ potential~\cite{Hertzberg:2017dkh, Byrnes:2018txb}.
In particular, proposed models in the literature specify the time-evolution of the second SR parameter, which is often re-defined as  
\begin{equation}
    \label{eqn: eta definition in terms of slow roll parameters}
    \eta(N) \equiv \epsilon_1(N) - \frac{1}{2}\epsilon_2(N)\;. 
\end{equation}
Note that for $\epsilon_1\ll 1$, $\eta$ is just a rescaling of $\epsilon_2$, and non-attractor behaviour corresponds to $\eta\gtrsim 3/2$. 
We label this approach to inflation in terms of $\eta(N)$ as a \textit{Hubble-flow parametrisation}.
The time-evolution of $\eta(N)$ could be obtained numerically from a potential model. Nevertheless, the Hubble-flow approach is especially useful when $\eta(N)$ is modeled by means of a simple analytical function --such as the one in Eq.~\eqref{eqn: eta parametrisation}--, whose shape resembles the $\eta(N)$ profiles observed in explicit microphysical models. This is the approach to Hubble-flow parametrisation most encountered in the literature.

Unlike conventional PBH-forming models based on an explicit potential formulation, specifying a simple analytical form for $\eta(N)$ offers several advantages. 
First, within this approach one appears to be able to study specific dynamics (e.g. transient non-attractor behavior) without committing to a single microphysical model.  
In this sense, such parametrisation seems to allow for model-independent studies.
Second, the $\eta(N)$ parametrisation gives more direct control over the inflationary dynamics. 
The value of $\epsilon_2$ during the non-attractor phase, its duration and the suddenness of the transitions in and out of it can all be adjusted independently, giving this approach considerable phenomenological flexibility and computational advantage~\cite{Ragavendra:2024yfp}.
For example, the $\eta(N)$ parametrisation gives immediate control over the maximum amplitude of $\mathcal{P}_\zeta$, which is determined by the integrated behaviour of $\eta(N)$, see Eq.~\eqref{eqn: zeta super-horizon solution1}.
On the other hand, controlling it through the inflaton potential requires the solution of background equations of motion. 
Moreover, fine adjustments of $V(\phi)$ are usually required to obtain the desired amplitude of $\mathcal{P}_\zeta$. 
Beyond linear-order, the dynamics of curvature perturbations are governed by higher-order interaction Hamiltonians, whose couplings are fixed by the slow-roll hierarchy. 
Parametrising inflation through functions such as $\eta(N)$ therefore provides a description of the time-dependent interactions entering the In--In formalism~\cite{Weinberg:2005vy}, allowing higher-point correlators and loop contributions to be computed without explicit reconstruction of the underlying inflaton potential.
For these reasons, Hubble-flow parametrisations have been widely employed in studies of PBH production from transient non-attractor phases~\cite{Ballesteros:2020sre, Ragavendra:2020sop, Franciolini:2022pav, Caravano:2024moy, Caravano:2025diq, Caravano:2026hca}, primordial non-Gaussianity~\cite{Taoso:2021uvl, Ballesteros:2024pbe, Ballesteros:2024pwn}, loop contributions to primordial correlators~\cite{Franciolini:2023agm, Ballesteros:2024zdp}, and the separate-universe approach~\cite{Raveendran:2025pnz}. 
See also Refs.~\cite{Byrnes:2018txb, Ballesteros:2020sre, Kristiano:2022maq, Riotto:2023hoz, Kristiano:2023scm, Motohashi:2023syh, Kristiano:2024vst} for $\eta(N)$ models featuring instantaneous transitions between the different dynamical phases, and Refs.~\cite{Cai:2018dkf, Firouzjahi:2023aum, Firouzjahi:2023ahg, Firouzjahi:2023bkt, Cruces:2026qvl} for alternative analytical methods to model sharp, instant and smooth transitions.

Our aim is to scrutinise the model-building of non-attractor inflation by comparing an analytical Hubble-flow parametrisation with analytical potentials.
Is one approach preferable to the other from a fundamental perspective? 
We show that, even for slowly-varying $\eta(N)$, the underlying potential responsible for the imposed dynamics hides sharp features in $V_{\phi\phi\phi}$ and higher derivatives, at the time of the transition from the non-attractor phase to the final attractor phase. 
Modeling inflation from $\eta(N)$ is not a model-independent procedure at all; the underlying potentials enforcing the required background dynamics include very specific localised features, leading to evolution of the perturbations quite different from that found in conventional models derived from smooth potentials.

\medskip
\textit{Content:}
This paper is organised as follows. 
In Sec.~\ref{sec: Modeling inflation from eta(N)}, we introduce a widely used Hubble-flow parametrisation for describing transient non-attractor models (Sec.~\ref{sec: hubble-flow parametrisation}) and derive the corresponding reconstructed potentials (Sec.~\ref{sec: reconstruction of the inflationary potential}). 
To investigate the impact of sharp features found in the potential on the evolution of the background and perturbations, in Sec.~\ref{sec: uv filtering} we implement a filtering procedure to suppress high-frequency modes of $V_{\phi\phi\phi}$ in field space, based on a discrete Fourier transform (DFT) approach. 
We apply the UV filter both to Hubble-flow-derived (Sec.~\ref{sec: Filtering eta(N) derived potentials.}) and analytical (Sec.~\ref{sec: UV Filtering analytical potentials models}) potentials, and discuss the implications of removing UV features at the level of the background evolution.
In Sec.~\ref{sec: Linear perturbation power spectrum} we look into the impact of UV features in Hubble-flow-derived potentials on linear perturbations, and their removal through the filter.
In particular, we discuss implications for Wands duality invariance~\cite{Wands:1998yp} in Sec.~\ref{subsec: Wands duality}, both for Hubble-flow-derived potentials (Sec.~\ref{sec: WDI eta models}) and analytical models (Sec.~\ref{sec: WDI potential models}). 
We present our conclusions in Sec.~\ref{sec: discussion}. 

\section{\texorpdfstring{Modeling inflation from $\bm{\eta(N)}$}{Modeling inflation from eta(N)}}
\label{sec: Modeling inflation from eta(N)}

We introduce a widely used Hubble-flow parametrisation scheme in Sec.~\ref{sec: hubble-flow parametrisation}, and describe how to reconstruct the corresponding inflationary potential in Sec.~\ref{sec: reconstruction of the inflationary potential}.

\subsection{Hubble-flow parametrisation}
\label{sec: hubble-flow parametrisation}

We consider the Hubble-flow parametrisation introduced in Ref.~\cite{Taoso:2021uvl} and used in Refs.~\cite{Caravano:2024moy, Caravano:2025diq, Caravano:2026hca} to describe non-attractor dynamics:
\begin{equation}
\label{eqn: eta parametrisation}
    \eta(N) = 
    \frac{1}{2} \left[\eta_{\mathrm{I}}+\eta_{\mathrm{III}}+(\eta_{\mathrm{II}}-\eta_{\mathrm{I}})\tanh\left(\frac{N - N_{\mathrm{in}}}{\delta N}\right) + (\eta_{\mathrm{III}} - \eta_{\mathrm{II}})\tanh\left(\frac{N -N_{\mathrm{out}}}{\delta N}\right)\right] \;.
\end{equation}
Here $N_{\rm in}$ and $N_{\rm out}$ approximately mark the transitions into and out of the non-attractor phase and $\delta N$ controls the corresponding rate of change of $\eta$. 
The parameters $\eta_{\rm I}$ and $\eta_{\rm III}$ respectively represent the values of $\eta(N)$ in the limits $N \ll N_{\rm in}$ and $N \gg N_{\rm out}$.  
For models with a  sufficiently-long intermediate phase, $\eta(N)\simeq\eta_{\rm II}$ when $N_{\rm in} \ll N \ll N_{\rm out}$. 
We work with the same parameter choices as in Refs.~\cite{Caravano:2024moy, Caravano:2025diq, Caravano:2026hca}, see Table~\ref{tab:USR_cases}. 
\begin{table}
\centering
\begin{tabular}{|c|c|c|c|c|c|}
    \hline
     &$\eta_{\rm I}$& $\eta_{\rm II}$ & $\eta_{\rm III}$ & $\Delta N \equiv N_{\rm out} - N_{\rm in}$ & $\delta N$ \\
    \hline
    Case $\mathbb{I}$ (Wands duality)&0 & 3.5 & -0.5 & $2.6 + 0.29 \log_{10} \mathcal{P}_\zeta^{\rm max}$ & 0.5 \\
    Case $\mathbb{II}$ (repulsive) &0& 3 & -0.5 & $3.3 + 0.38 \log_{10} \mathcal{P}_\zeta^{\rm max}$ & 0.5 \\
    Case $\mathbb{III}$ (attractive)&0 & 4.5 & -0.5 & $1.8 + 0.19 \log_{10} \mathcal{P}_\zeta^{\rm max}$ & 0.5 \\
    \hline
\end{tabular}
\caption{Values of the parameters in Eq.~\eqref{eqn: eta parametrisation} for the three realisations of Hubble-flow parametrisations considered in Refs.~\cite{Caravano:2024moy, Caravano:2025diq,Caravano:2026hca}.}
\label{tab:USR_cases}
\end{table}
In particular, $\eta_{\rm I} = 0$ ensures that CMB scales exited the horizon during slow roll, and $\eta_{\rm III}<0$ leads to the end of inflation. 
The parameter choice $\delta N=0.5$ models a slow transition between the different regimes, and the duration of the non-attractor phase is chosen such that the maximum of the power spectrum is $\mathcal{P}_{\zeta}^{\rm max}=10^{-2}$.  

The three cases specified in Table~\ref{tab:USR_cases} are labeled according to the sign of $V_{\phi \phi \phi}$ near the second transition, i.e. the transition out of the non-attractor phase. 
Cases $\mathbb{II}$ and $\mathbb{III}$ correspond to models with $V_{\phi \phi \phi}$ positive (\textit{repulsive}) and negative (\textit{attractive}) respectively~\cite{Caravano:2024moy, Caravano:2025diq, Caravano:2026hca}. 
Case $\mathbb{I}$, referred to as \textit{Wands duality}, has the smallest $V_{\phi \phi \phi}$. 
We elaborate on implications of Hubble-flow parametrisation for Wands duality in Sec.~\ref{subsec: Wands duality}.

\begin{figure}
    \centering
    \includegraphics[width=1.0\linewidth]{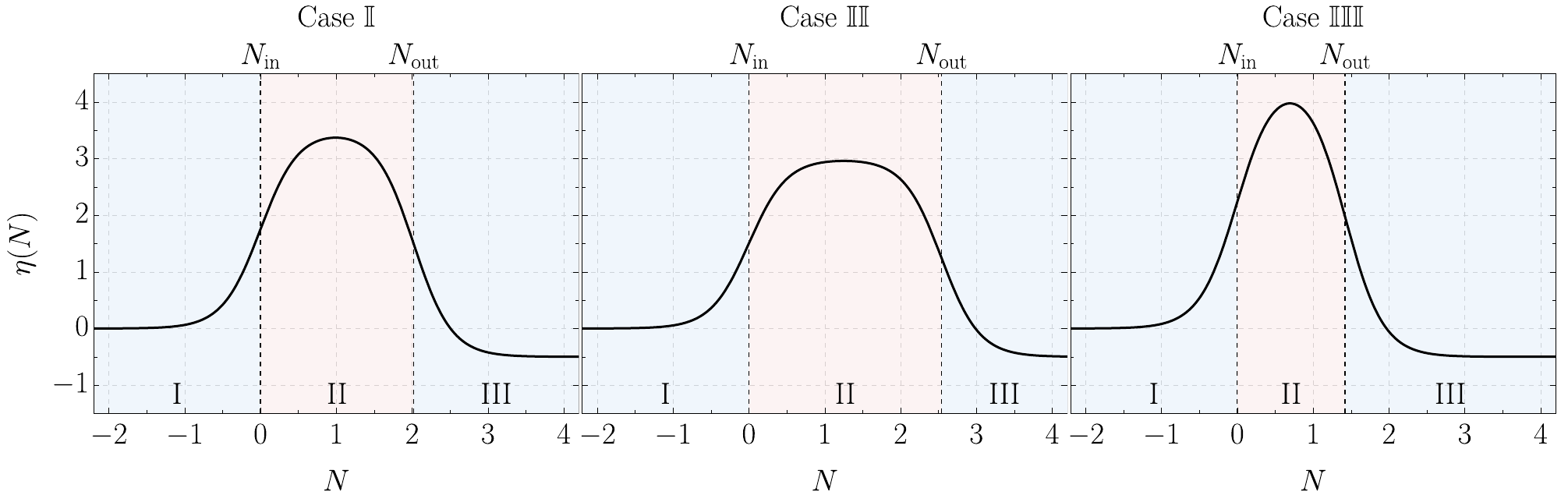}
    \caption{Evolution of $\eta(N)$ as a function on the number of e-folds, $N$, normalised such that $N_{\rm in}=0$. 
    Each panel corresponds to one of the three models considered in Refs.~\cite{Caravano:2024moy, Caravano:2025diq,Caravano:2026hca}, see Table~\ref{tab:USR_cases}. 
    Region I and III are characterised by attractor dynamics. 
    Region II corresponds to the non-attractor phase, and it is defined by $\eta(N) \gtrsim 3/2$.   
    Vertical black dashed lines mark $N_{\rm in}$ and $N_{\rm out}$.}
    \label{fig: etas original eta model}
\end{figure}

In Fig.~\ref{fig: etas original eta model} we show the time-evolution of $\eta(N)$ for the three models listed in Table~\ref{tab:USR_cases}. 
The non-attractor phase is realised when $\eta(N) \gtrsim 3/2$.
For $\epsilon_1\ll 1$, this leads to super-horizon growth of $\zeta$, see Eq.~\eqref{eqn: zeta super-horizon solution1}. 
Note that the transitions in and out of the non-attractor phase do not necessarily happen exactly at $N_{\rm in}$ and $N_{\rm out}$. 
However, these values provide a good approximation for the transition times.

\subsection{Reconstruction of the inflationary potential}
\label{sec: reconstruction of the inflationary potential}

During inflation, the evolution of the background is dictated by the Klein--Gordon and Raychaudhuri equations
\begin{equation}
    \label{eqn: eom backgorund scalar field}
    \ddot{\phi}+3H \dot{\phi}+V_{\phi}=0 \quad \text{and} \quad {\dot H} = -\frac{1}{2} {\dot \phi^2}  \;,
\end{equation}
under the Friedmann constraint 
\begin{equation}
	H^2 = \frac{1}{3} \left[ \frac{1}{2} \dot \phi^2 + V(\phi) \right] \;.
\end{equation}
Here we set $M_{\text{Pl}}=1$. 
To reconstruct the potential required to realise a specific $\eta(N)$ parametrisation~\cite{Byrnes:2018txb, Franciolini:2022pav}, such as the one in Eq.~\eqref{eqn: eta parametrisation}, it is useful to rewrite the Klein--Gordon and Friedmann equations in terms of the independent variables $V(N)$ and $\epsilon_1(N)$,  
\begin{equation}
    \label{eqn: eom background scalar field in terms of e-folds and epsilon_1}
    \frac{\mathrm{d} \epsilon_1}{\mathrm{d}N} + \left(2\epsilon_1 + \frac{\mathrm{d}\ln V}{\mathrm{d}N}\right)(3-\epsilon_1)=0 
    \quad \text{and} \quad
     V = H^2(3-\epsilon_1) \;,
\end{equation}
where we have switched from cosmic time to e-folds. 
We can write the Raychaudhuri equation (\ref{eqn: eom backgorund scalar field}) as
\begin{equation}
\label{eqn: epsilon_1 in terms of phi'}
    \epsilon_1(N)
    = 
    \frac{1}{2} 
    \,
    \left( {\frac{\mathrm{d}\phi}{\mathrm{d}N}} 
    \right)^2 \;. 
\end{equation}
The solution for $\epsilon_1(N)$ corresponding to a given $\eta(N)$ parametrisation can be obtained by solving numerically Eq.~\eqref{eqn: eta definition in terms of slow roll parameters}, which is nothing but a differential equation for $\epsilon_1(N)$
 \begin{equation}
 \label{eq:depsilon1bydN}
    \eta(N) = \epsilon_1(N) - \frac{1}{2}\frac{\mathrm{d} \ln \epsilon_1(N)}{\mathrm{d}N} \;. 
\end{equation}
With it, Eq.~\eqref{eqn: eom background scalar field in terms of e-folds and epsilon_1} can then be solved for $V(N)$, yielding
\begin{equation}
\label{eq: V(N) solution}
    V(N) = V(N_{\text{CMB}}) \exp \left\{ -2 \int_{N_{\text{CMB}}}^{N} \mathrm{d}N' \left[ \frac{\epsilon_1(3 - \eta)}{3 - \epsilon_1} \right] \right\} \;,  
\end{equation}
where we have provided boundary conditions at $N_{\rm CMB}$, i.e. when the CMB pivot scale exited the horizon.   
To reconstruct the potential in field space, $V(\phi)$, one has to exchange e-folds for field values. 
The definition of $\epsilon_1(N)$, Eq.~\eqref{eqn: epsilon_1 in terms of phi'}, yields
\begin{equation}
    \label{eqn: phi(N)}
    \phi(N) = \phi(N_{\text{CMB}}) \pm \int_{N_{\text{CMB}}}^{N} \mathrm{d}N' \sqrt{2\epsilon_1} \;. 
\end{equation}
For monotonic $\phi(N)$, this can be inverted to express $N$ in terms of $\phi$. 
By substituting the resulting $N(\phi)$ in Eq.~\eqref{eq: V(N) solution}, one obtains $V(\phi)$.
Without loss of generality, we choose in Eq.~\eqref{eqn: phi(N)} the negative sign, corresponding to $\phi$ decreasing with $N$. 
 
\begin{figure}
    \centering
    \includegraphics[width=1.0\linewidth]{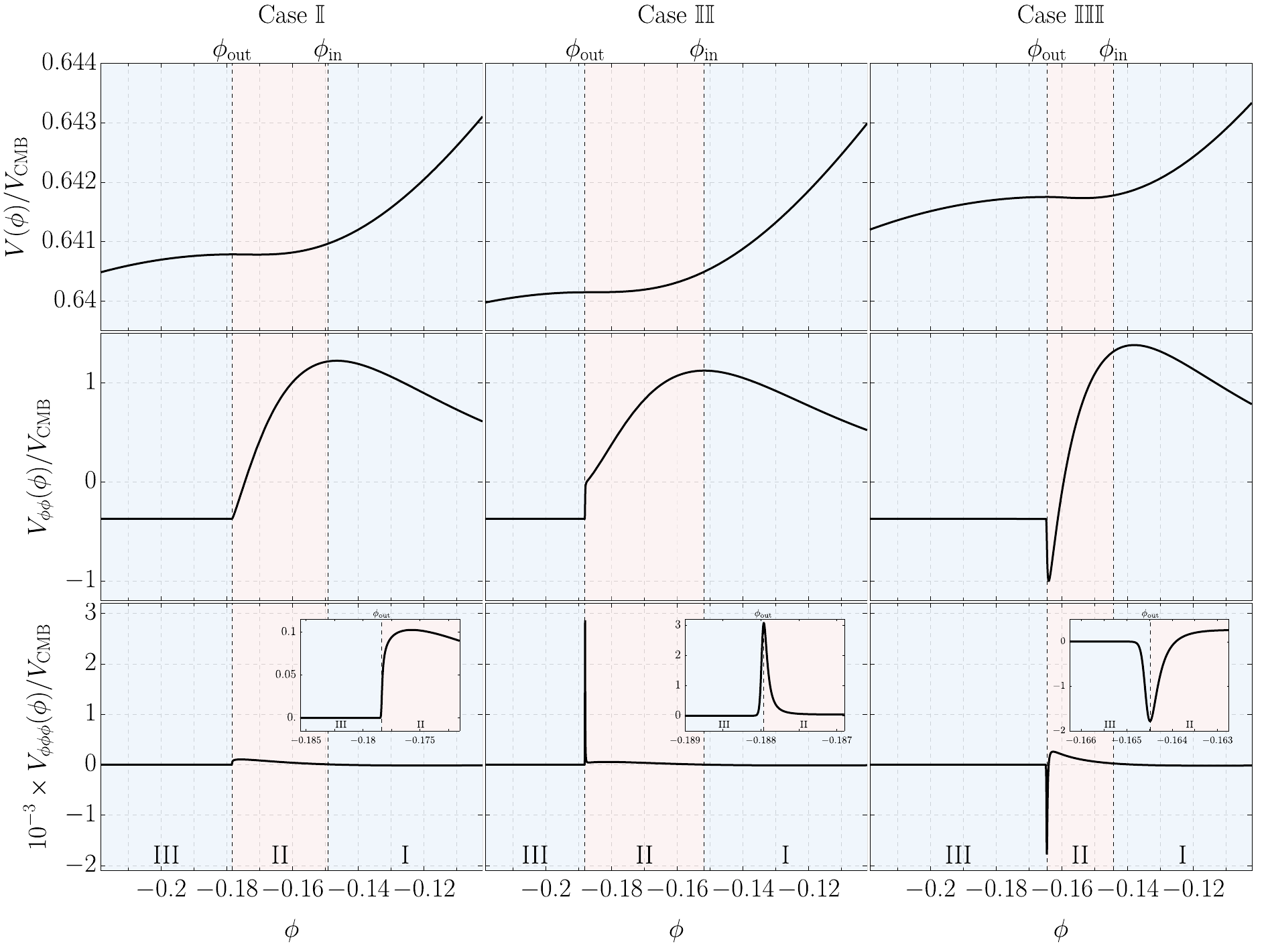}
    \caption{Reconstructed inflationary potentials $V(\phi)$ for each one of the $\eta(N)$ models considered in Refs.~\cite{Caravano:2024moy, Caravano:2025diq,Caravano:2026hca}, see Sec.~\ref{sec: hubble-flow parametrisation}.
    From top to bottom, we display the inflationary potential and its second and third derivatives, normalized by $V_{\rm CMB}$. 
    For visualization purposes the third derivative is further rescaled by a factor of $10^{-3}$.
    In the bottom panels, the inset plot shows a zoom-in of the sharp feature in $V_{\phi\phi\phi}$ at the time of the second transition.
    Vertical black, dashed lines mark the field values $\phi_{\rm in} \equiv \phi(N_{\rm in})$ and $\phi_{\rm out} \equiv \phi(N_{\rm out})$. 
    }
    \label{fig:potentials original eta model}
\end{figure}

With this procedure, we reconstruct the potentials necessary to yield the $\eta(N)$ evolution specified by Eq.(\ref{eqn: eta parametrisation}) for cases $\mathbb{I}$, $\mathbb{II}$ and $\mathbb{III}$. 
We impose initial conditions for the field and its velocity at $N_{\rm CMB} = -30$, in particular $\phi(N_{\rm CMB}) = 3.5$ and $\mathrm{d}\phi/\mathrm{d}N |_{N_{\rm CMB}} = -0.11$~\footnote{Note that the reconstructed potential depends on the initial conditions. 
For example, for different initial field velocities the corresponding $\epsilon_1$ solutions in Eq.~\eqref{eq:depsilon1bydN} differ by an overall amplitude $\epsilon_1(N_{\rm CMB})$. The field excursion $\Delta \phi$, see Eq.~\eqref{eqn: discrete field sample}, is stretched for larger initial field velocities, while preserving the shape of $V(\phi)$. Therefore, the corresponding $V_{\phi\phi\phi}$ display an excess of UV power (see Sec.~\ref{sec: uv filtering}) independent on the choice of the initial field velocity.}, and normalise the potential such that $V_{\rm CMB}\equiv V(N_{\rm CMB}) = 3.2 \times 10^{-9}$~\cite{Caravano:2024moy, Caravano:2025diq,Caravano:2026hca}.
The reconstructed potentials and their derivatives are represented in Fig.~\ref{fig:potentials original eta model}, where we have zoomed in on the region corresponding to the non-attractor phase.
We see that in all three cases, $V_{\phi\phi}$ increases smoothly as the field approaches the first transition into the non-attractor phase, reaching a maximum value at $\phi\sim\phi_{\rm in}$.
Although the potentials look very smooth, cases $\mathbb{II}$ and $\mathbb{III}$ have step-like features in $V_{\phi\phi}$ near the second transition, i.e. at $N_{\rm out}$. 
This results in sharp, localised spikes in $V_{\phi\phi\phi}$.  
Note that we consider slow transitions in $\eta(N)$; for a more abrupt variation of $\eta(N)$, i.e. for $\delta N \to 0$, the features in $V_{\phi\phi\phi}$ are further enhanced.
For Case $\mathbb{I}$, engineered to satisfy Wands duality invariance (see Sec.~\ref{subsec: Wands duality}), $V_{\phi\phi\phi}$ is approximately one order of magnitude smaller than in cases $\mathbb{II}$ and $\mathbb{III}$, and exhibits a step rather than a spike.

The appearance of sharp features in $V_{\phi\phi\phi}$ even in the presence of slow variation of $\eta(N)$ (see Fig.~\ref{fig: etas original eta model}) can be understood in terms of the time-dependence of $\phi(N)$ during the non-attractor phase.
When $\eta\gtrsim1$, $\epsilon_1$ decreases exponentially with respect to $N$ (see Eq.~\eqref{eqn: eta definition in terms of slow roll parameters} and Fig.~\ref{fig: eta and eps1 filtered eta models}), and $\phi$ becomes almost constant in $N$. 
Therefore slow variation over several e-folds in $V(N)$ becomes compressed into rapid variation in $V(\phi)$ almost at a single value of $\phi$.

 \section[\texorpdfstring{Characterising and filtering UV power in $\bm{V(\phi)}$}{Characterising and filtering UV power in V}]{Characterising and filtering UV power in $\bm{V(\phi)}$}
\label{sec: uv filtering}

In Sec.~\ref{sec: Filtering eta(N) derived potentials.} we will use a discrete Fourier transform (DFT) to characterise the sharp features in $V_{\phi\phi\phi}$, seen in Sec.~\ref{sec: reconstruction of the inflationary potential} in the reconstructed potential at the transition out of the non-attractor phase.
We then introduce a Fourier low-pass filtering procedure, which allows us to suppress UV modes in field space.
By applying the UV filter to the inflationary potential and its derivatives, we show the effect that a progressive removal of UV power has on the background evolution for $\eta(N)$ models.

In Sec.~\ref{sec: UV Filtering analytical potentials models}, we compare Hubble-flow parametrisations against the explicit $V(\phi)$ models reviewed in Ref.~\cite{Cole:2023wyx}.
By applying an identical processing and filtering pipeline to both cases, we treat all models on an equal footing and systematic effects due to the methodology cancel in relative comparisons.

Before discussing their application in Secs.~\ref{sec: Filtering eta(N) derived potentials.} and~\ref{sec: UV Filtering analytical potentials models}, let us describe the DFT and UV filter. 
Ultimately we will apply them to $V(\phi)$ \textit{and} its derivatives, but for illustration we present here the case of $V(\phi)$. 
By defining the field range over which inflation occurs, $\Delta \phi \equiv \phi_{\rm end}-\phi_{\rm start}$, we discretise field values as
\begin{equation}
\label{eqn: discrete field sample}
    \phi_j = \phi_{\rm start} + j\,\delta\phi \;, 
    \quad
    \text{where}
    \quad 
    j = 0,\, 1,\, \ldots,\, n 
    \quad 
    \text{and}
    \quad 
    \delta\phi = \frac{\Delta \phi}{n} \;. 
\end{equation}
Here $\phi_{\rm start}$ and $\phi_{\rm end}$ denote the values of the inflaton at the beginning and end of our numerical solutions.
A discrete sample of the potential is then given by the array of length $n+1$,
\begin{equation}
\label{eqn: discrete sample potential}
    [V_0, \, V_1, \, \ldots,\, V_n],
    \quad 
    \text{where}
    \quad
    V_j \equiv V(\phi_j) \;.
\end{equation} 
The uniform spacing between adjacent sampling points, $\delta \phi$, determines the resolution with which the potential $V(\phi)$ is discretised prior to applying the DFT. 
For fixed $\Delta \phi$, this is set by the total number of samples, $n$. 
We fix $n$ to ensure that all relevant features of the potential are adequately captured, allowing for a reliable reconstruction of $\mathcal{P}_\zeta(k)$. 
In particular, $n$ is chosen such that the difference between $\mathcal{P}_\zeta$ computed from the original potential and that obtained from the sampled potential remains below $0.1\%$. This ensures that any appreciable discrepancies in subsequent analyses are driven by physical effects rather than discretisation bias.

The DFT assumes periodicity of the data being transformed. 
In order to avoid the Gibbs phenomenon~\cite{1899Natur..59..606G}, an oscillatory artefact that arises near discontinuities when a function is represented by a filtered Fourier series, we extend the dataset~\eqref{eqn: discrete sample potential} by constructing a symmetric array of length $2n + 1$, 
\begin{equation}
    \mathcal{V} =  \left[V_n,\, V_{n-1},\,\ldots,\,V_1,\,V_0,\,V_1,\,\ldots,\,V_{n-1},\,V_n \right]\;.
\end{equation} 
We label the single element of this array as $\mathcal{V}_\ell$, where $\ell = 0,\, 1,\, \ldots,\, 2n$. 
The DFT of the potential is then defined as 
\begin{equation}
    \label{eqn: V DFT}
    \tilde V = [\tilde{V}_{k_{\phi_0}},\, \ldots,
    \tilde{V}_{k_{\phi_{2n}}}], 
    \quad 
    \text{where}
    \quad
    \tilde{V}_{k_{\phi_m}} = \sum_{\ell=0}^{2n} \mathcal{V}_{\ell} \, e^{-i k_{\phi_m} \phi_\ell}\;, 
\end{equation}
where $m=0,\,1,\,\ldots,\,2n$, $k_{\phi_m} \equiv 2\pi m /[(2n+1)\delta \phi]$ is the Fourier wavenumber and $\phi_\ell = \ell \delta \phi$. Here $\phi_\ell$ is an indexed coordinate defined for the DFT procedure only, and does not represent discrete field values. 

By means of a suitable window function, it is possible to attenuate high-frequency components while preserving the low-frequency ones. 
We choose to employ a Gaussian-type window function,
\begin{equation}
    W_{k_{\phi,c}}= \exp \left(-\frac{k_{\phi_m}^2}{k_{\phi,c}^2}\right) \;, 
\end{equation}
which yields the filtered potential in field domain
\begin{multline}
    \label{eqn: V DFT and window function}
    V^{\text{filt}}(k_{\phi,c}) = [V^{\text{filt}}_0(k_{\phi,c}),\, \ldots,\, 
    V^{\text{filt}}_{2n}(k_{\phi,c})],
    \\
    \text{where}
    \quad 
    V^{\text{filt}}_\ell(k_{\phi,c}) 
    =
    \frac{1}{2n+1} \sum_{m=0}^{2n}
    \tilde{V}_{k_{\phi_m}}\, W_{k_{\phi,c}}\, e^{i k_{\phi_m} \phi_\ell} \;.
\end{multline}
Here $V^{\text{filt}}(k_{\phi,c})$ is a smoothed version of the original potential in which high-frequency features ($k_\phi\gtrsim k_{\phi,c}$) have been suppressed. 
The cut-off wavenumber $k_{\phi,c}$ defines the minimum field-space excursion over which features in the potential can be resolved after filtering,
\begin{equation}
    \label{eqn: delta phi_c definition}
    \delta\phi_c \equiv \frac{2\pi}{k_{\phi,c}} \;. 
\end{equation}
The Gaussian window function is real and enjoys the same even-symmetry as the potential dataset.
Therefore, the filtering procedure yields real and even-symmetric $V^{\text{filt}}_\ell(k_{\phi,c})$. 
Consequently, the first $n$ values ($\ell=0,\ldots,n-1$) are the mirrored extension of the original data and are entirely determined by the remaining points. We discard the mirrored extension and retain the values corresponding to the original physical field range.

In Fourier analysis, the asymptotic decay of Fourier coefficients is controlled by the regularity class of the function being transformed; for any continuous function whose $m$-th derivative is piecewise continuous the Fourier coefficients decay as $k^{-(m+1)}$ for $k\to \infty$. 
An even-mirrored dataset has a first derivative that is piecewise continuous, therefore its Fourier coefficients decay as $k_{\phi}^{-2}$.
Here our aim is to characterise UV features of $V_{\phi\phi\phi}$, and therefore we must choose whether to sample and even-mirror $V$ or $V_{\phi\phi\phi}$. 
One could sample $V_{\phi\phi\phi}$, even-mirror the resulting dataset and then apply the DFT (\textit{method A}). 
This procedure has the advantage of attenuating the intrinsic UV scaling of the power spectrum of $V_{\phi\phi\phi}$, as we shall demonstrate below.
Alternatively, one could even-mirror the potential $V$, apply the DFT, and then obtain the Fourier coefficients of $V_{\phi\phi\phi}$ by deriving those of the potential
\footnote{The (continuum) Fourier transform of the derivative $V_\phi$ is $\tilde{V}_{\phi_{k_\phi}} = \int \mathrm{d}\phi\, V_{\phi}\,e^{-i k_{\phi} \phi}$. Integrating by parts and assuming that boundary terms vanish due to periodicity/mirroring construction  yields $\tilde{V}_{\phi_{k_\phi}}
= ik_{\phi}\,\tilde{V}_{k_\phi}$.} 
(\textit{method B}). 
The resulting Fourier coefficients scale as ${\tilde{V}_{\phi\phi\phi}} \propto k_\phi^3 \, \tilde V_{k_\phi} \propto k_{\phi}$, where the $k_{\phi}^3$ factor is due to differentiation and we have used the aforementioned scaling for an even-mirrored dataset. 
This implies a power spectrum that grows in the UV, $k_{\phi}|\tilde{V}_{\phi \phi \phi}|^2\propto k_{\phi}^3$. 
Within \textit{method A}, we first compute $V_{\phi \phi \phi}$ in field space and then apply the mirroring procedure, yielding $\tilde{V}_{\phi \phi \phi}\propto k_\phi^{-2}$.
In this case, the DFT power spectrum decays in the UV, $k_{\phi}|\tilde{V}_{\phi \phi \phi}|^2\propto k_{\phi}^{-3}$.
We choose to apply \textit{method A}, as the removal of UV power due to the interplay of discontinuities and Fourier transforms allows us to highlight the genuine UV features in the spectrum of ${V}_{\phi \phi \phi}$.  

\subsection{The case of Hubble-flow-derived potentials}
\label{sec: Filtering eta(N) derived potentials.}

Here we apply the DFT and UV filter to the reconstructed potentials presented in Sec.~\ref{sec: reconstruction of the inflationary potential}.
For all cases, we fix the number of samples $n=5\times10^6$.
The field excursion is $\Delta \phi \simeq 5.58$, with differences of order $\mathcal{O}(10^{-2})$ between the three different models, and therefore the spacing between adjacent sampling points $\delta \phi$, see Eq.~\eqref{eqn: discrete field sample}, remains effectively unchanged. 

\begin{figure}
    \centering
    \includegraphics[width=1.0\linewidth]{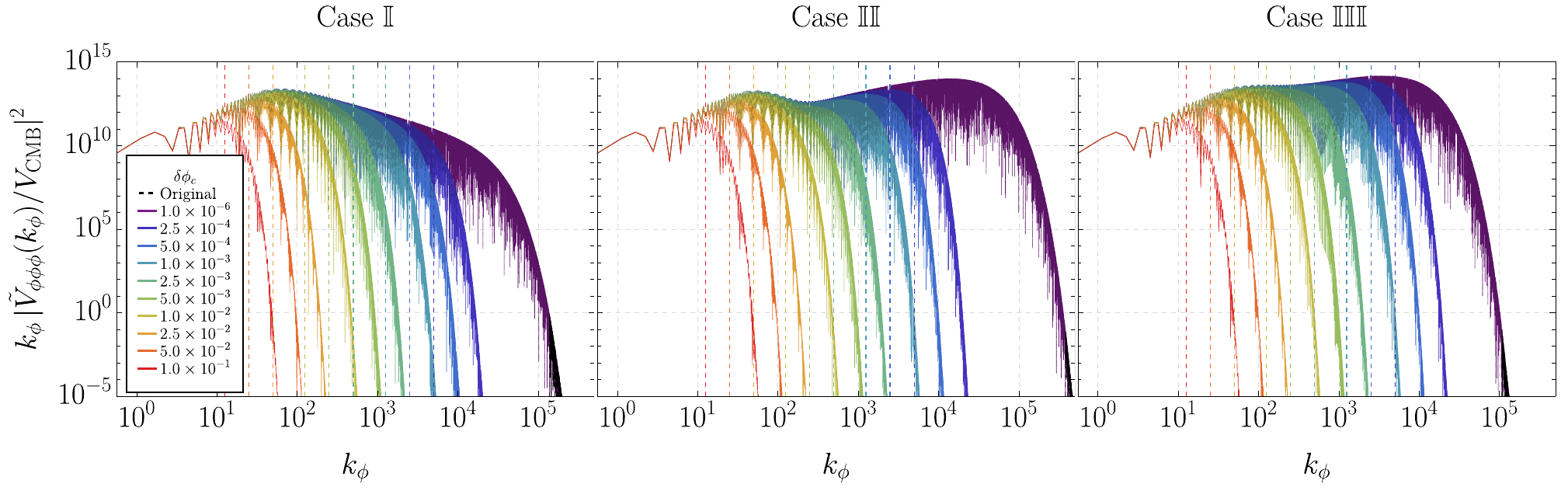}
    \caption{
    Power spectrum of $V_{\phi\phi\phi}$ for the reconstructed inflationary potentials of Sec.~\ref{sec: reconstruction of the inflationary potential}, represented as a function of wavenumber $k_{\phi}$. 
    The original spectrum (dashed, black line) is naturally truncated at the maximum wavenumber set by the discrete field-space sampling of $V(\phi)$ underlying the DFT.
    The colored lines represent filtered spectra obtained for different cut-off wavenumber, $k_{\phi,c}$, see Eq.~\eqref{eqn: V DFT and window function}.
    Values of $k_{\phi,c}$ are marked by vertical, dashed lines. 
    In the legend, we connect $k_{\phi,c}$ to the corresponding cut-off resolution in field space, $\delta \phi_c$, see Eq.~\eqref{eqn: delta phi_c definition}.
    In each panel, the spectrum of the original potential (represented with a dashed, black line) is barely visible as it is almost the same as the spectrum obtained with the first UV filter (purple line).}
    \label{fig: V''' power filtered eta(N) models}
\end{figure}

The DFT power spectrum for $V_{\phi\phi\phi}$ for each of the three models given in Table~\ref{tab:USR_cases} is displayed in Fig.~\ref{fig: V''' power filtered eta(N) models}.
By inspecting the spectra of the original potentials, one sees that they all display the same blue tilt up to $k_{\phi} \sim 10^2$. 
However, for larger $k_\phi$ cases $\mathbb{II}$ and $\mathbb{III}$ display a distinctive bump of UV power, while case $\mathbb{I}$ turns into a red-tilted spectrum.
This is consistent with the sharp features observed in the bottom panel of Fig.~\ref{fig:potentials original eta model}.
The characteristic width of the sharp feature in case $\mathbb{II}$ is smaller than in Case $\mathbb{III}$, which corresponds to a spectrum with support further into the UV.
The colored lines are produced by applying the UV filter~\eqref{eqn: V DFT and window function} to the original spectrum. 
The filtered spectrum with highest resolution closely reproduces all features of the original spectrum.
All other filtered spectra are defined by evenly spaced cut-off wavenumbers, $k_{\phi,c}$, and in each of them high-frequency power $(k_\phi\gtrsim   k_{\phi,c})$ is suppressed. 

\begin{figure}
    \centering
    \includegraphics[width=1.0\linewidth]{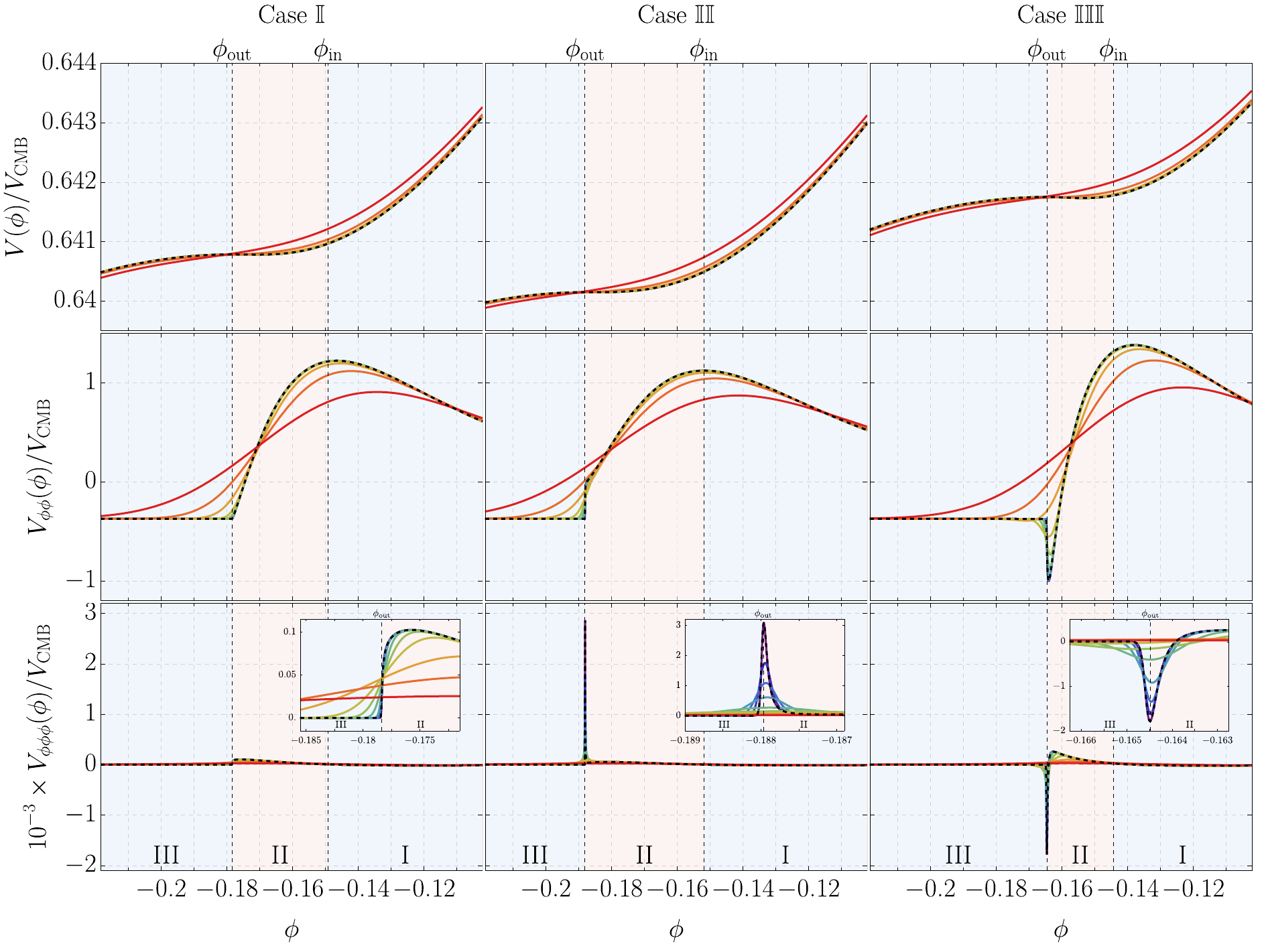}
    \caption{Reconstructed potentials from Sec.~\ref{sec: reconstruction of the inflationary potential} and their higher-order derivatives, both in their original form (black, dashed line) and after UV filtering (colored lines), see Eq.~\eqref{eqn: V DFT and window function}. 
    In the bottom panels, the inset plot shows a zoom-in of the sharp feature in $V_{\phi\phi\phi}$ at the time of the second transition.
    See Fig.~\ref{fig: V''' power filtered eta(N) models} for the color legend indicating the filter cut-off wavenumber, $k_{\phi,c}$.}
    \label{fig: filtered potentials eta model}
\end{figure}

In Fig~\ref{fig: filtered potentials eta model} we show the effect that the UV filter has on the potential and its derivatives $V_{\phi\phi}$ and $V_{\phi\phi\phi}$. 
Interestingly, at the level of the potential nearly all filters produce very similar results, with all curves lying almost exactly on top of the original black, dashed line. 
However, the sharp features observed in the derivatives are progressively smoothed out by the UV filter, eventually leading to a nearly flat $V_{\phi\phi\phi}$. 
The UV filter primarily affects the second transition.  
For example, by inspecting the $V_{\phi\phi}$ panels for cases $\mathbb{II}$ and $\mathbb{III}$, one sees that for the smallest $k_{\phi,c}$ the step-like feature corresponding to the second transition has been completely smoothed away, while the broad bump corresponding to the first transition is reduced but still present. 

\begin{figure}
    \centering
    \includegraphics[width=1.0\linewidth]{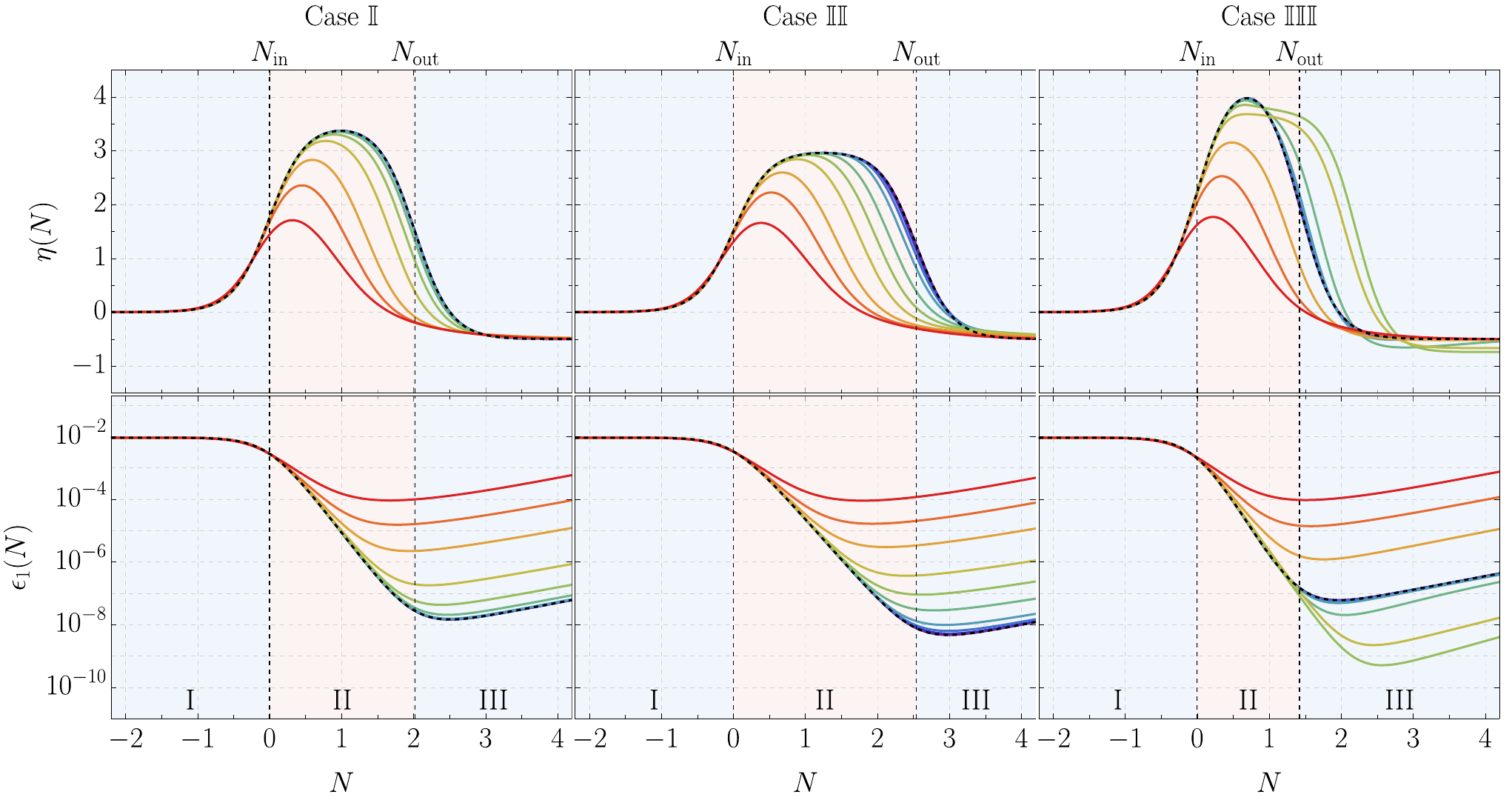}
    \caption{Time-evolution of $\eta(N)$ (top panels) and $\epsilon_1(N)$ (bottom panels), both in their original form (black, dashed line) and after UV filtering (colored lines).
    The dashed, black lines in the top panels are the same as in Fig.~\ref{fig: etas original eta model}. 
    See Fig.~\ref{fig: V''' power filtered eta(N) models} for the color legend indicating the filter cut-off wavenumber, $k_{\phi,c}$.}
    \label{fig: eta and eps1 filtered eta models}
\end{figure}

In order to assess the impact that the progressive reduction of UV power has on the background evolution, we solve the background dynamics by using the UV filtered potentials. 
Results for $\eta(N)$ and $\epsilon_1(N)$ are displayed in Fig.~\ref{fig: eta and eps1 filtered eta models}. 
As expected, we recover the pre-imposed $\eta(N)$ dynamics for the nearly unfiltered potentials. 
For the other filters, the new $\eta(N)$ profiles generally are bounded by the envelope of the original profile, with $\eta(N)$ peaking at smaller values.
This is in turn reflected in $\epsilon_1(N)$, which progressively decreases less and less during the non-attractor phase. 
Moreover, the end of inflation $(\epsilon_1=1)$ is typically reached earlier with the filtered potentials, with the exception of some of the intermediate filters (shown in green) in Case $\mathbb{III}$. 
Overall, we observe that the background dynamics is very sensitive to UV filtering. 
This is expected as the filter disrupts the delicate fine-tuning of the potential that is necessary in the first place to deliver sustained non-attractor dynamics.  

\subsection{The case of analytical potentials}
\label{sec: UV Filtering analytical potentials models}

In contrast to the Hubble-flow-parametrisation approach, inflationary dynamics are usually solved starting from a specific microphysical model, $V(\phi)$. 
As done in Sec.~\ref{sec: Filtering eta(N) derived potentials.} for the $\eta(N)$-reconstructed potentials, our aim here is to characterise the UV power of prototypical analytical potentials leading to a transient non-attractor phase, and gauge the effect that the UV filter has on them. 
In particular, we consider the four analytical potentials discussed in Ref.~\cite{Cole:2023wyx}, which are representative of the typical mechanisms leading to inflaton deceleration, the hallmark of non-attractor dynamics. 
We introduce each potential below, and report in Table~\ref{tab: analytical potential parameters} the fiducial parameters used in each case. 
\begin{table*}
\centering
\renewcommand{\arraystretch}{1.3}
\setlength{\tabcolsep}{7pt}
\scriptsize

\begin{tabular}{|ll|ll|ll|ll|}
\hline

\multicolumn{2}{|c|}{\textbf{Non-exponential}}
    & \multicolumn{2}{c|}{\textbf{Exponential}}
    & \multicolumn{2}{c|}{\textbf{Polynomial}}
    & \multicolumn{2}{c|}{\textbf{Gaussian bump}} \\
\hline

      Param. & Value
    & Param. & Value
    & Param. & Value
    & Param. & Value \\
\hline

$\phi_{\rm start}$ & $3$
    & $\phi_{\rm start}$ & $12$
    & $\phi_{\rm start}$ & $-|\Lambda^3/c_3|$
    & $\phi_{\rm start}$ & $4$ \\

$\phi_{\rm CMB}$ & $2.3909$
    & $\phi_{\rm CMB}$ & $9.26703$
    & $\phi_{\rm CMB}$ & $-0.0015312$
    & $\phi_{\rm CMB}$ & $3$ \\

$a$ & $1/\sqrt{2}$
    & $V_0$ & $1.714\times 10^{-9}$
    & $V_0$ & $1.540\times 10^{-14}$
    & $V_0$ & $7.247\times10^{-11}$ \\

$b$ & $3/2$
    & $A_W$ & $2/100$
    & $\Lambda$ & $0.3$
    & $n$ & $2$ \\

$\lambda$ & $1.86\times10^{-6}$
    & $B_W$ & $1$
    & $c_0$ & $1$
    & $M$ & $1/2$ \\

$v$ & $0.19669$
    & $C_W$ & $4/100$
    & $c_1$ & $\frac{2\pi^2\Lambda^4}{4225c_3}$
    & $A$ & $1.17\times10^{-3}$ \\

& 
    & $\langle\tau_{K3}\rangle$ & $14.30$
    & $c_2$ & $0$
    & $\sigma$ & $1.59\times10^{-2}$ \\

&
    & $G_W/\langle\mathcal{V}\rangle$ & $3.08054\times10^{-5}$
    & $c_3$ & $-0.52$
    & $\phi_d$ & $2.18812$ \\

&
    & $R_W/\langle\mathcal{V}\rangle$ & $7.071067\times10^{-4}$
    & $c_4$ & $1$
    & & \\

&
    &
    & 
    & $c_5$ & $-0.640725043$
    & & \\

\hline
\end{tabular}
\normalsize
\caption{Parameter choices for the four analytical potentials analysed in Sec.~\ref{sec: UV Filtering analytical potentials models}.}
\label{tab: analytical potential parameters}
\end{table*}
Note that introducing a PBH-producing short-scale peak in $\mathcal{P}_\zeta$ is sometimes in tension with CMB measurements on large scales (see e.g. Ref.~\cite{Iacconi:2021ltm}). 
However, in this work we are not concerned with the observational viability of the potentials discussed. 

\paragraph{Deceleration via non-polynomial potential feature -- Non-exponential character.}
We consider an inflection-point model\footnote{An inflection point is generated for $b = 1 - \frac{a^2}{3} + \frac{a^2}{3}\left(\frac{9}{2a^2} - 1\right)^{2/3}$.} inspired by Higgs inflation~\cite{Hamada:2014iga, Bezrukov:2014bra}, in which the Higgs field is non-minimally coupled to gravity via $\mathcal{L} \supset \xi h^2 R$. 
Upon conformal transformation to the Einstein frame and canonical normalisation of the Higgs, the Einstein-frame potential displays a flat region at large field values, where slow-roll inflation can take place~\cite{Bezrukov:2014bra}.
By inserting an inflection point at intermediate field values, it is possible to realise a transient non-attractor phase. 
We consider the potential first proposed in~\cite{Garcia-Bellido:2017mdw} and later adapted in~\cite{Germani:2017bcs}, 
\begin{equation}
    V(\phi) = \frac{\lambda}{12}\,\phi^2 v^2\,
    \frac{\displaystyle 6 - 4a\,\frac{\phi}{v} + 3\,\frac{\phi^2}{v^2}}
         {\displaystyle\left(1 + b\,\frac{\phi^2}{v^2}\right)^2}.
    \label{eqn: non-exp character potential}
\end{equation}

\paragraph{Deceleration via non-polynomial potential feature -- Exponential character.}
The second potential we consider also displays an inflection point at intermediate field values, after the CMB scales crossed the horizon. 
However, in this case the potential is motivated by string theory~\cite{Cicoli:2018asa}, and the inflaton deceleration is due to exponential factors. 
Moreover, an effective shift symmetry of the inflaton at large $\phi$ guarantees stability against quantum corrections. 
The model is defined as
\begin{equation}
    V(\phi) = V_0 \Bigg[ C_1 
    - e^{-\sqrt{\frac{1}{3}}\,\hat{\phi}}
    \left(
    1 - \frac{C_6}{1 - C_7 e^{-\sqrt{\frac{1}{3}}\,\hat{\phi}}}
    \right) + C_8 e^{\sqrt{\frac{2}{3}}\,\hat{\phi}}
    \left(
    1 - \frac{C_9}{1 + C_{10} e^{\sqrt{3}\,\hat{\phi}}}
    \right)
    \Bigg],
    \label{eqn: exp character potential}
    \end{equation}
    where $\hat \phi = \phi -\sqrt{3}/2 \,\ln\langle \tau_{K3}\rangle$ and the potential parameters relate to the parameters of the underlying string construction as
    \begin{equation}
    C_6 = \frac{A_W}{C_W} \;,  
    \quad 
    C_7 = \frac{B_W}{\langle \tau_{K3} \rangle^{1/2}} \;, 
    \quad 
    C_8 = 0 \;, 
    \quad
    C_8 C_9 = \frac{G_W}{\langle \mathcal{V} \rangle}
    \frac{\langle \tau_{K3} \rangle^{3/2}}{C_W} \;, 
    \quad
    C_{10} = \frac{R_W}{\langle \mathcal{V} \rangle}
    \langle \tau_{K3} \rangle^{3/2} \;, 
\end{equation}
with $C_1$ chosen such that $V_{\min} = 0$. 

\paragraph{Deceleration via polynomial potential feature.}
In the third potential we consider the inflaton deceleration is achieved by an inflection point in a polynomial potential. 
In particular, Ref.~\cite{Hertzberg:2017dkh} tunes a quintic potential to have two almost-flat regions, one during which the CMB scales crossed the horizon and one leading to a transient non-attractor phase. 
The potential reads 
\begin{equation}
    V(\phi) = V_0\left(c_0 - \frac{c_1}{\Lambda}\phi + \frac{c_2}{2\Lambda^2}\phi^2 - \frac{c_3}{3!\Lambda^3}\phi^3 + \frac{c_4}{4!\Lambda^4}\phi^4 - \frac{c_5}{5!\Lambda^5}\phi^5\right),
    \label{eqn:polynomial_feature_potential}
\end{equation}
where $c_i$, $\Lambda$, and $V_0$ are free parameters. 
Note that we have reversed the signs of the odd-power terms in the original potential of~\cite{Hertzberg:2017dkh}, so that the inflaton rolls leftwards. 
Apart from requiring fine-tuned parameters for PBH production, this model is also sensitive to initial conditions, requiring the inflaton velocity to be very close to zero.

\paragraph{Deceleration via superposed feature -- Gaussian bump.}
Finally, we consider a potential where the inflaton deceleration is obtained from a localised Gaussian bump (or dip), added \textit{ad hoc} on top of a potential that would otherwise sustain slow-roll inflation. 
The realisation we consider was put forward in Ref.~\cite{Mishra:2019pzq}, 
\begin{equation}
    V(\phi) = V_0 \frac{\phi^n}{\phi^n + M^n}
    \left(1 + A \exp\left[-\frac{(\phi - \phi_d)^2}{2\sigma^2}\right]\right),
    \label{eqn: gaussian bump potential}
\end{equation}
where $A$, $\phi_d$ and $\sigma$ characterise the height, position and  width of the Gaussian bump, and the underlying KKLT slow-roll potential~\cite{Kallosh:2019eeu} is described by the parameters $V_0$, $M$ and $n$.
Clearly the potential~\eqref{eqn: gaussian bump potential} allows one to separate small-field dynamics from the large-scale ones, thereby removing the tension between PBH production and CMB observations. 
On the other hand, due to its \textit{ad hoc} construction, it seems to be less motivated by an underlying physical model with respect to the others we consider. 

\medskip
For each potential, we solve the background equations of motion starting from $N_{\rm start}=0$ up to the end of inflation. 
We list in Table~\ref{tab: analytical potential parameters} the inflaton initial condition, $\phi(N_{\rm start})\equiv \phi_{\rm start}$, and set  $\mathrm{d}\phi/\mathrm{d}N|_{N_{\rm start}} = 0$. 
The field excursion during inflation, $\Delta \phi$, varies amongst the four potentials considered, from $\Delta \phi \sim 1$ for the polynomial model to $\Delta \phi \sim 10$ for the exponential model. 
This implies that for fixed $n$ the sampling resolution, see Eq.~\eqref{eqn: discrete field sample}, will vary. 
In order to provide a fair comparison against the $\eta(N)$ models, for each potential we choose $n$ such that the sampling resolution is the same as for the $\eta(N)$ models, see Fig.~\ref{fig: V''' power filtered eta(N) models}. 

\begin{figure}
    \centering
    \includegraphics[width=1.0\linewidth]{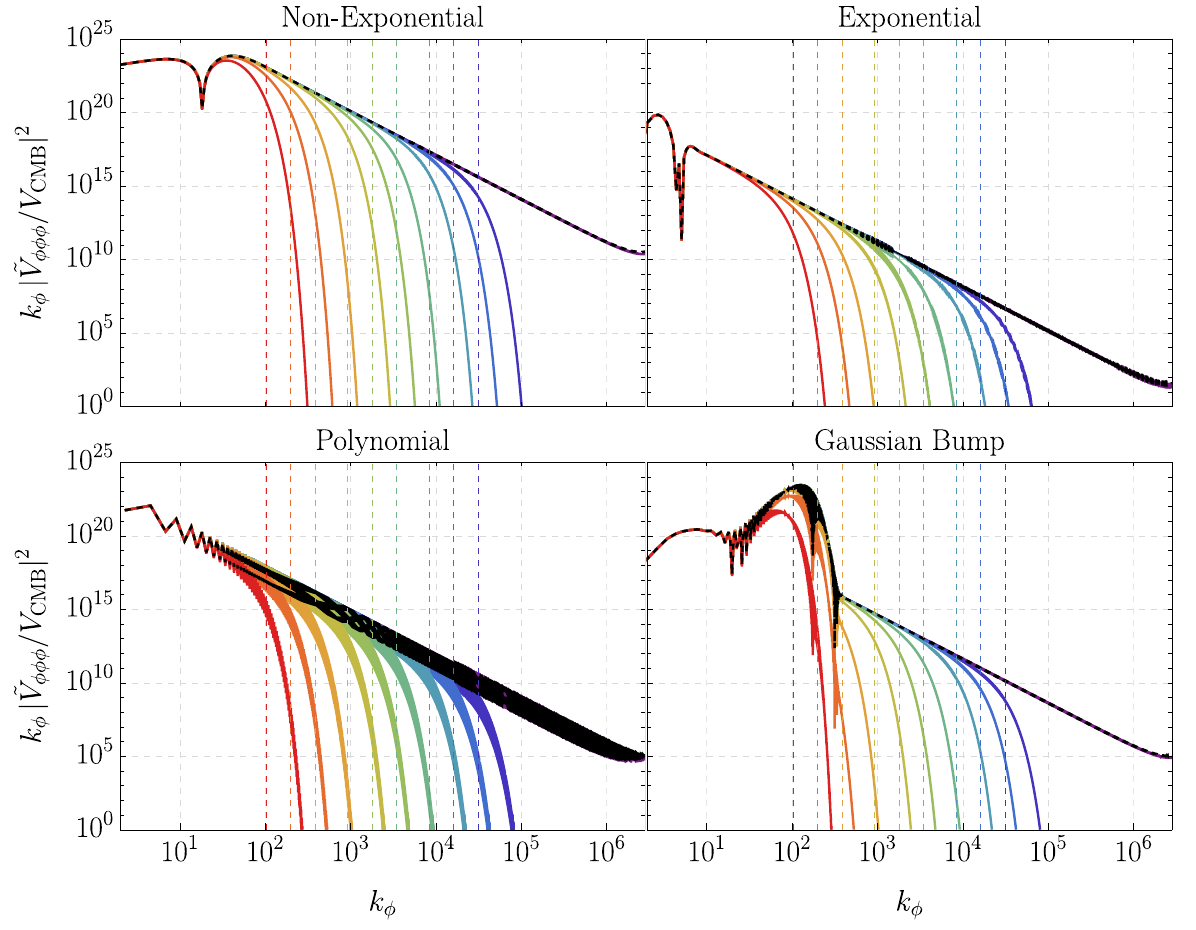}
    \caption{Power spectrum of $V_{\phi\phi\phi}$ for the four analytical potentials analysed in Sec.~\ref{sec: UV Filtering analytical potentials models}, represented as a function of wavenumber $k_\phi$. 
    The original spectrum is represented with a dashed, black line, while the colored ones represent filtered spectra obtained for different cut-off wavenumber, $k_{\phi,c}$, see Eq.~\eqref{eqn: V DFT and window function}. 
    Values of $k_{\phi,c}$ are marked by vertical, dashed lines.  
    See Fig.~\ref{fig: V''' power filtered eta(N) models} for the color legend indicating the minimum field-space excursion over which features can be resolved after filtering, $\delta\phi_{c}$.
    }
    \label{fig: dft pwer spectrum analytical potentials}
\end{figure}

In Fig.~\ref{fig: dft pwer spectrum analytical potentials} we display the power spectrum of $V_{\phi\phi\phi}$ for the potentials~\eqref{eqn: non-exp character potential}, ~\eqref{eqn: exp character potential}, ~\eqref{eqn:polynomial_feature_potential} and~\eqref{eqn: gaussian bump potential}. 
In contrast to those of the reconstructed potentials shown in Fig.~\ref{fig: V''' power filtered eta(N) models}, we observe a red-tilted spectrum $\propto k_{\phi}^{-3}$ in the UV for all analytical potentials. 
In this case the spectrum is dominated by the natural asymptotic decay arising from the DFT procedure, see below Eq.~\eqref{eqn: delta phi_c definition}. 
Unlike the spectra in Fig.~\ref{fig: V''' power filtered eta(N) models}, there is no excess of UV power corresponding to sharp features in field space. 
All differences in the spectra are confined to low frequencies. 
The onset of the power-law regime occurs at $k_{\phi} \sim 10^2$ for the non-exponential and at $k_{\phi} \sim 10^1$ for the exponential model.
The polynomial model exhibits power-law behaviour extending to the lowest frequencies.
The Gaussian bump model is the only one with a localised excess in power at intermediate frequencies, peaking at $k_{\phi}\sim 10^2$, before the UV power-law tail takes over from $k_{\phi}\sim 5\times 10^2$. 

\begin{figure}
    \centering
    \includegraphics[width=1.0\linewidth]{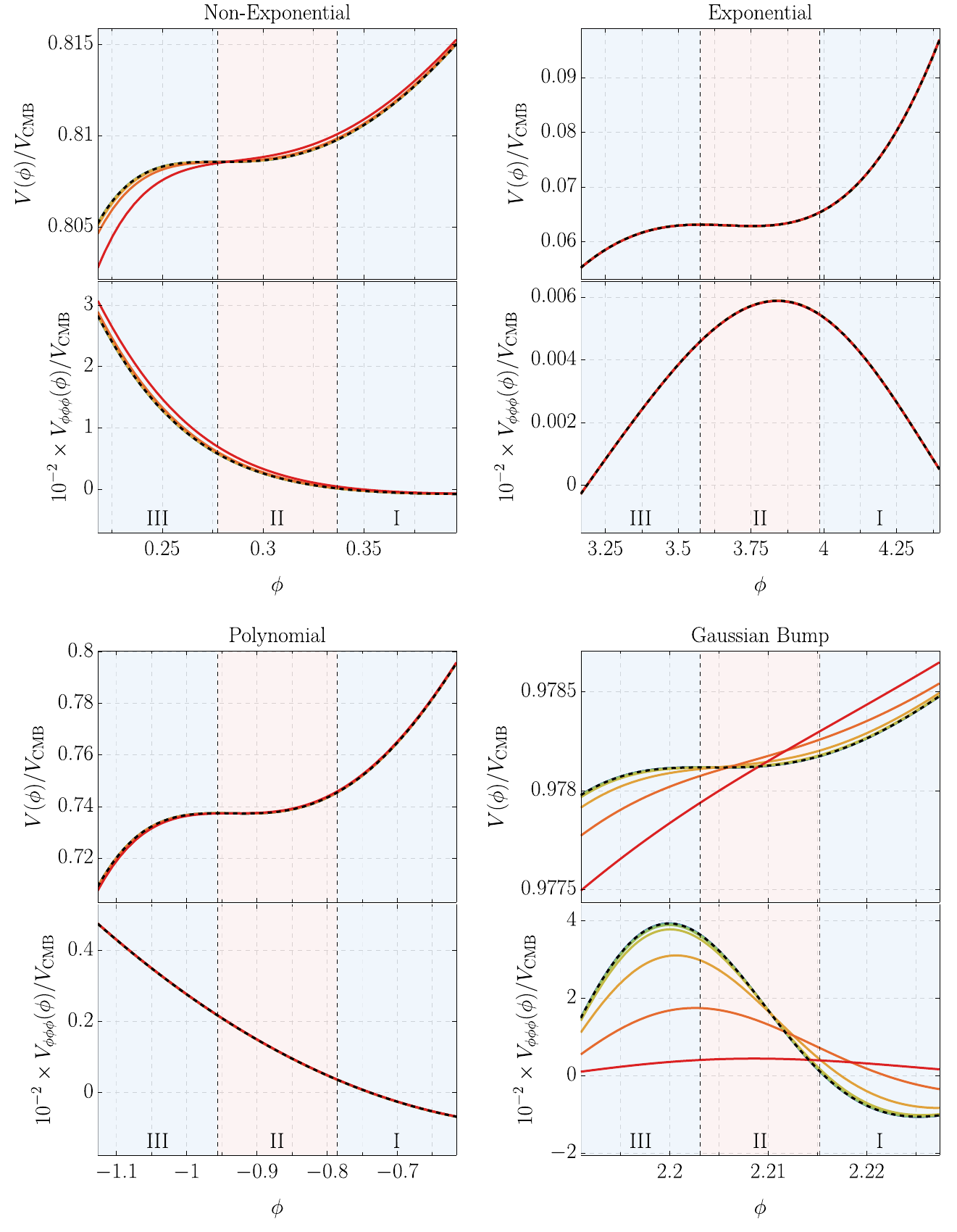}
    \caption{Analytical potentials presented in Sec.~\ref{sec: UV Filtering analytical potentials models} and their higher-order derivative $V_{\phi\phi\phi}$, both in their original form (black, dashed line) and after UV filtering (colored lines), see Eq.~\eqref{eqn: V DFT and window function}. 
    See Fig.~\ref{fig: V''' power filtered eta(N) models} for the color legend indicating the minimum field-space excursion over which features can be resolved after filtering, $\delta\phi_{c}$. The red shaded region corresponds to the non-attractor phase, $\eta \gtrsim 3/2$, of the unfiltered model. 
    }
    \label{fig: analytical potentials and third derivative}
\end{figure}

The spectra displayed in Fig.~\ref{fig: dft pwer spectrum analytical potentials} reflect the field-space shape of $V_{\phi\phi\phi}$ shown in Fig.~\ref{fig: analytical potentials and third derivative}. 
Unlike the reconstructed potentials discussed in Sec.~\ref{sec: Modeling inflation from eta(N)}, for all analytical models $V_{\phi\phi\phi}$ 
is smooth and does not exhibit sharp features. 
We observe that the filtered $V(\phi)$ and $V_{\phi\phi\phi}(\phi)$ are almost indistinguishable from the original ones, with the exception of the Gaussian bump potential. 

\begin{figure}
    \centering
    \includegraphics[width=1.0\linewidth]{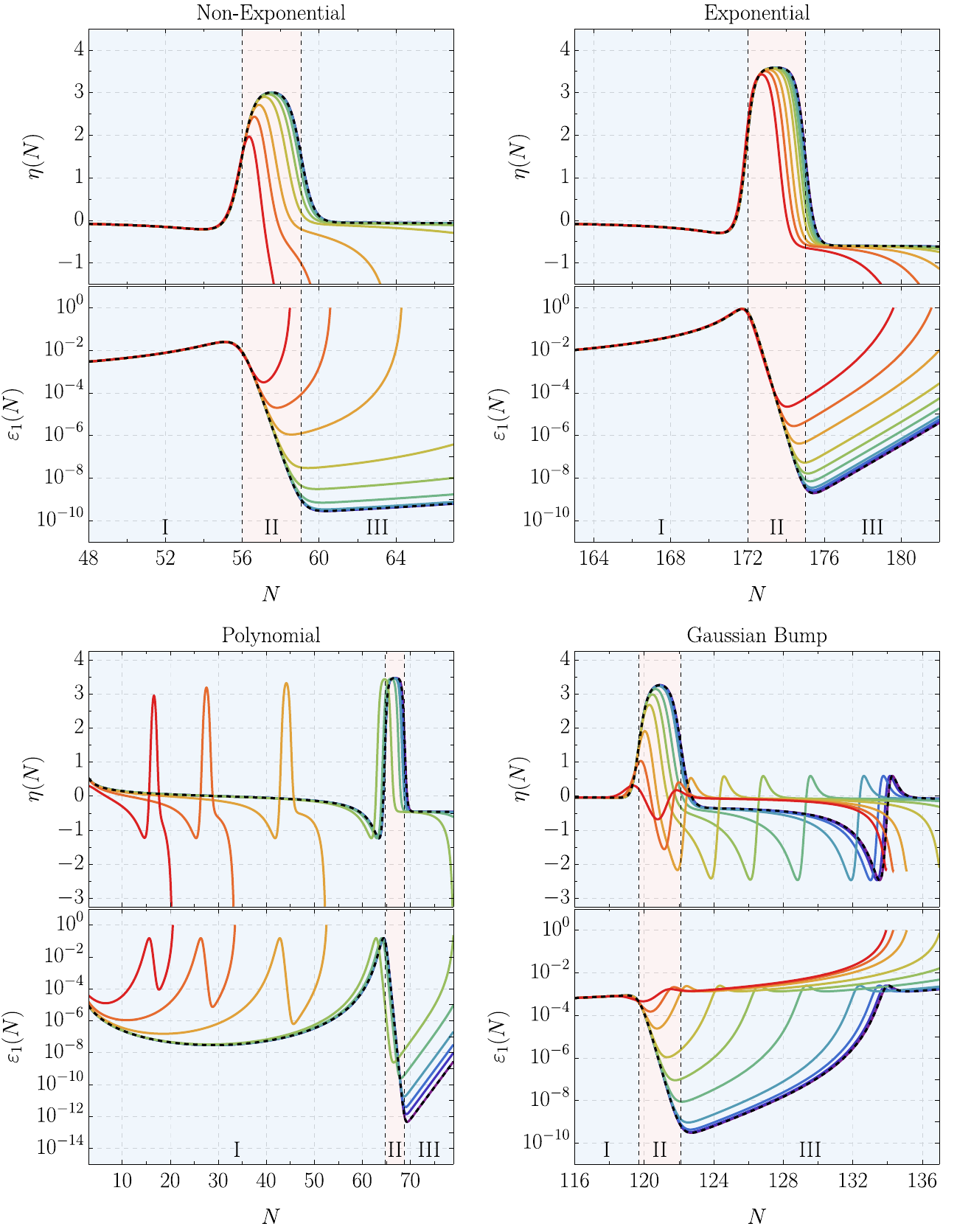}
    \caption{Time-evolution of $\eta(N)$ (top panels) and $\epsilon_1(N)$ (bottom panels), both in their original form (black, dashed line) and after UV filtering (colored lines).
    Each quadrant corresponds to one of the models presented in Sec.~\ref{sec: UV Filtering analytical potentials models}.
    See Fig.~\ref{fig: V''' power filtered eta(N) models} for the color legend indicating the minimum field-space excursion over which features can be resolved after filtering, $\delta\phi_{c}$. The red shaded region corresponds to the non-attractor phase, $\eta \gtrsim 3/2$, of the unfiltered model.}
    \label{fig: eta and eps1 filtered analytical potentials}
\end{figure}

In Fig.~\ref{fig: eta and eps1 filtered analytical potentials} we show the time-evolution of the first two slow-roll parameters obtained from the analytical potentials before and after UV filtering. 
With the exception of the Gaussian bump potential, the UV filter only induced small changes in the potentials, see Fig.~\ref{fig: analytical potentials and third derivative}. 
Nevertheless, the background dynamics are affected by it. 
Both the duration of the non-attractor phase and the value of $\eta$ during it are reduced. 
In addition, the polynomial model exhibit a distinct behaviour, with the first transition being shifted towards earlier times.
In the Gaussian bump model both phases of deceleration of the inflaton are affected in their magnitude and onset. 
Across all models, the filtering procedure renders the potential less flat, thereby causing inflation to end earlier. 
As seen for the Hubble-flow reconstructed potentials, the sensitivity of the background evolution to the UV filter again reflects the fine tuning of models leading to PBH formation. 

Our results show that prototypical analytical potentials leading to PBH formation are not characterised by sharp features in their higher-order derivatives, in contrast to potentials reconstructed from the analytical $\eta(N)$ parametrisation in Eq.~\eqref{eqn: eta parametrisation}. 
Analytical potentials do not require spikes in their higher derivatives because the USR phase is transient, and naturally relaxes back to the final attractor phase in the absence of features, whereas the reconstructed potentials are required to reproduce the specified form of the second transition in $\eta(N)$ from USR back to the attractor.

\section{Linear scalar perturbations}
\label{sec: Linear perturbation power spectrum}

We have so far explored the UV spectrum of both $\eta(N)$-reconstructed and analytical potentials and assessed the impact that removal of UV power has on the background dynamics. 
We now move on to consider linear perturbations. 
In Sec.~\ref{sec: power spectrum computation} we give a brief recap on their evolution and on the numerical procedure followed to compute the dimensionless spectrum $\mathcal{P}_\zeta(k)$. 
In Sec.~\ref{subsec: Wands duality} we establish whether the evolution respects so-called Wands duality invariance~\cite{Wands:1998yp, Kinney:2005vj, Tzirakis:2007bf, Morse:2018kda}. 
This is the invariance of perturbations under homogeneous background transformations, in this context between the non-attractor phase and following attractor era. 
We make an explicit connection between duality invariance and the behavior of $V_{\phi\phi\phi}$ at the second transition. 
We then assess the impact that the progressive removal of UV power has on Wands duality invariance and the spectrum of perturbations. 
Results for Hubble-flow models are presented in Sec.~\ref{sec: WDI eta models} and for analytical potentials in Sec.~\ref{sec: WDI potential models}. 

\subsection{Perturbations dynamics and power spectrum computation}
\label{sec: power spectrum computation}

In conformal time, $\mathrm{d}\tau \equiv \mathrm{d}t/a(t)$, the equation of motion for the linear comoving curvature perturbation, $\zeta_k(\tau)$, is
\begin{equation}
    \label{eqn: eom zeta conformal time}
    \zeta_k''(\tau) + \frac{z'(\tau)}{z(\tau)}\zeta_k'(\tau)+k^2 \zeta_k(\tau) =0 \;, 
\end{equation}
where $z\equiv a \dot{\phi}/H$ and a prime indicates a derivative with respect to $\tau$.
By introducing the Mukhanov--Sasaki variable $v_k \equiv z\, \zeta_k$, Eq.~\eqref{eqn: eom zeta conformal time} yields~\cite{Sasaki:1986hm, Mukhanov:1988jd}
\begin{equation}
    \label{eqn: MS equation}
    v_k''(\tau) + \left[k^2-\frac{z''(\tau)}{z(\tau)}\right]v_k(\tau)=0 \;. 
\end{equation}
Here the time-dependent mass term, usually referred to as the Mukhanov--Sasaki mass, takes the following exact form in terms of SR parameters
\begin{equation}
    \label{eqn: m^2 in terms of slow roll parameters}
    \mu^2\equiv \frac{1}{(aH)^2}\frac{z''}{z}
    = 2 - \epsilon_1
    + \frac{3}{2}\,\epsilon_{2}
    + \frac{1}{4}\,\epsilon_{2}^{2}
    - \frac{1}{2}\,\epsilon_1\,\epsilon_{2}
    + \frac{1}{2}\,\epsilon_{2}\,\epsilon_{3} \;. 
\end{equation}
We compute the linear power spectrum of $\zeta_k$ by solving Eq.~\eqref{eqn: MS equation} in terms of the number of e-folds, $N$. 
We solve separately for the real and imaginary parts of $v_k$. 
For each mode we initialise the differential solver at $N_i$, chosen to be 5 e-folds before horizon crossing $(k=aH)$, and impose Bunch--Davies initial conditions at $N_i$, 
\begin{equation}
\Re(v_k)=\frac{1}{\sqrt{2k}}\;,
\quad 
\Im(v_k)=0 \;,
\quad 
\Re\left(\frac{\mathrm{d}v_k}{\mathrm{d}N}\right)=0\;, 
\quad 
\Im\left(\frac{\mathrm{d}v_k}{\mathrm{d}N}\right)=-\frac{\sqrt{k}}{\sqrt{2}\,k_i} \;, 
\end{equation}
where $k_i \equiv a(N_i)H(N_i)$. 
We evolve each mode until the end of inflation, at e-folding time $N_\text{end}$. 
The dimensionless curvature power spectrum is then computed as
\begin{equation}
\mathcal{P}_{\zeta}(k)=\frac{k^3}{2\pi^2}\left|\frac{v_k(N_\text{end})}{z(N_\text{end})}\right|^2 \;. 
\end{equation}

\subsection{Wands duality invariance}
\label{subsec: Wands duality}
Wands duality invariance (WDI) holds when the perturbation $v_k$ evolves in the same way in different backgrounds. 
Eq.~\eqref{eqn: MS equation} dictates the evolution of $v_k$, and WDI is therefore realised between backgrounds that leave the Mukhanov--Sasaki mass unaltered.  
In order to understand under which conditions WDI is satisfied, it is useful to write the Mukhanov--Sasaki mass, Eq.~\eqref{eqn: m^2 in terms of slow roll parameters}, in terms of potential derivatives
\begin{equation}
    \label{eqn: MS mass}
    \mu^2=2+5\epsilon_1-2\epsilon_1^2+2\epsilon_1\epsilon_2-\frac{V_{\phi\phi}}{H^2} \;, 
\end{equation}
where 
\begin{equation}
    \label{eqn: V'' in terms of slow roll parameters}
    \frac{V_{\phi\phi}}{H^2}=6\epsilon_1-\frac{3}{2}\epsilon_2-2\epsilon_1^2-\frac{1}{4}\epsilon_2^2+\frac{5}{2}\epsilon_1\epsilon_2-\frac{1}{2}\epsilon_2\epsilon_3 \;. 
\end{equation}
As previously noted, during USR the first slow-roll parameter decays exponentially (see Eq~\eqref{eqn: eta definition in terms of slow roll parameters} and Figs.~\ref{fig: eta and eps1 filtered eta models} and \ref{fig: eta and eps1 filtered analytical potentials}) and we can work in the decoupling limit ($\epsilon_1 \to 0$), in which case Eq.~\eqref{eqn: MS mass} reduces to
\begin{equation}
    \label{eqn: m^2 decoupling limit}
    \lim_{\epsilon_1 \to 0} \mu^2 = \left(2-\frac{V_{\phi \phi}}{H^2}\right) \;. 
\end{equation} 
In models with a transient non-attractor phase, WDI may be realised between the non-attractor and subsequent attractor era if the Mukhanov--Sasaki mass remains constant during these phases \textit{and} during the transition between them.
In the decoupling limit, this corresponds to 
\begin{equation}
    \label{eqn: m^2 time variation}
    \lim_{\epsilon_1 \to 0}\frac{\mathrm{d} \mu^2}{\mathrm{d}N} \propto \lim_{\epsilon_1 \to 0}\sqrt{\epsilon_1}\,V_{\phi\phi\phi}\to 0 \;,
\end{equation}
where we have used Eq.~\eqref{eqn: m^2 decoupling limit} and $H=\text{const}$. 
WDI invariance is therefore realised provided that $V_{\phi\phi\phi}$ is small enough such that the condition~\eqref{eqn: m^2 time variation} is satisfied at all times from the non-attractor phase into the subsequent attractor era.

For models that realise WDI, the backgrounds during the non-attractor and final attractor phases are related by a homogeneous transformation that leaves $V_{\phi\phi}$ (and hence the MS mass) invariant. 
For phases with constant $\epsilon_2$  $(\epsilon_3=0)$
\footnote{Equivalently, one can obtain the background transformation by solving Eq.~\eqref{eqn: V'' in terms of slow roll parameters} for constant $\epsilon_2$, which yields $\epsilon_2 =-3 \pm \sqrt{9-4V_{\phi \phi}/H^2}$, where the two solutions are related by $\epsilon_2^+ +\epsilon_2^- = -6$.}, this gives~\cite{Atal:2018neu, Ng:2021hll, Karam:2022nym, Briaud:2025hra}
\begin{equation}
    \label{eqn: background transf WDI}
    \epsilon_2 \to -\epsilon_2-6 \quad \text{or} \quad \eta \to -\eta+3 \;.
\end{equation}
For example, the dual of ultra-slow roll $(\epsilon_2=-6)$ is slow-roll $(\epsilon_2=0)$, while for generic non-attractor constant roll $(\epsilon_2<-6)$ the dual attractor phase has $\epsilon_2>0$. 

In Sec.~\ref{sec: WDI eta models} we discuss WDI in the context of Hubble-flow models and of analytical potentials in Sec.~\ref{sec: WDI potential models}. 

\subsubsection{The case of Hubble-flow-derived potentials}
\label{sec: WDI eta models}

\begin{figure}
    \centering
    \includegraphics[width=1.0\linewidth]{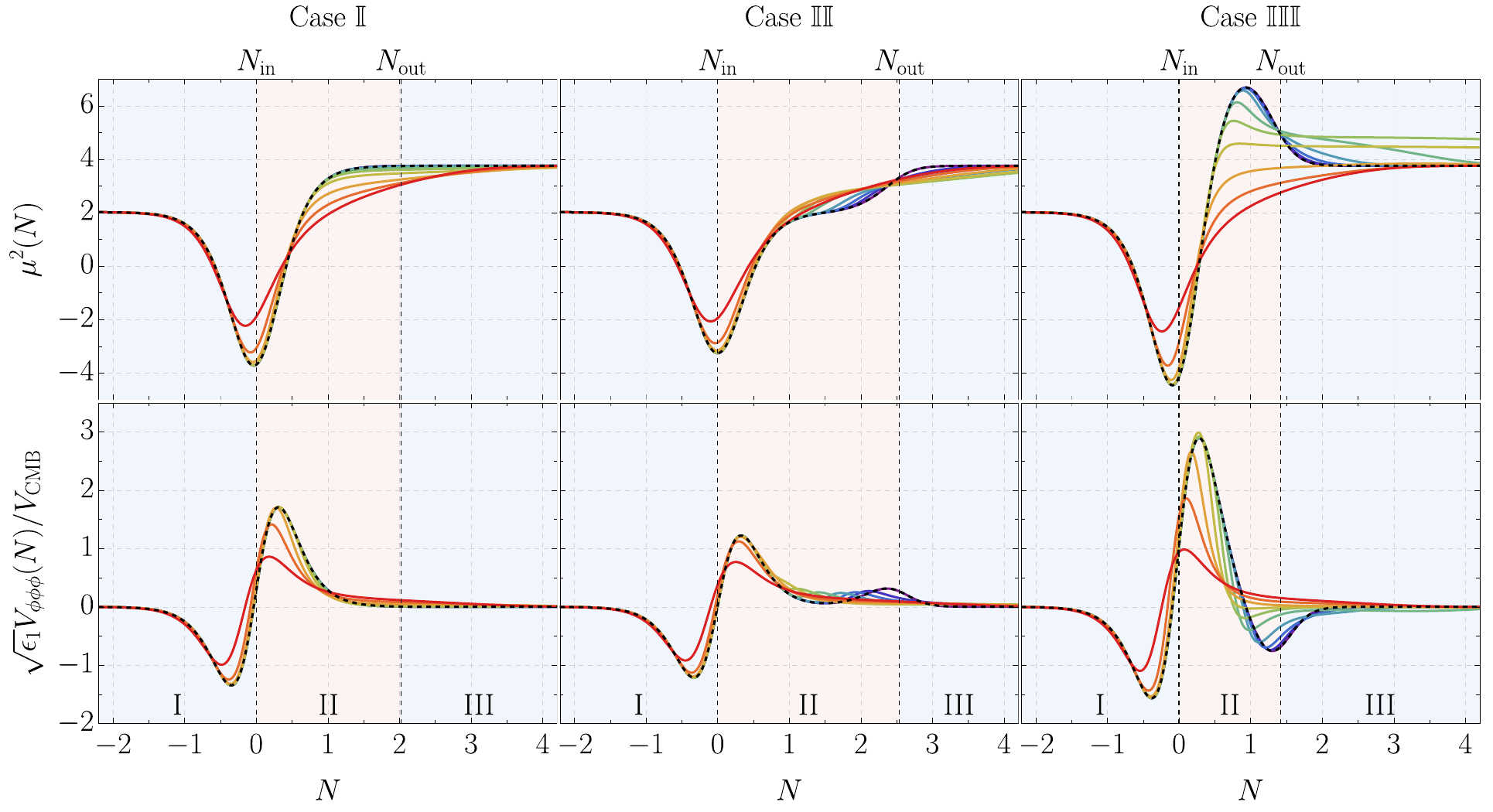}
    \caption{\textit{Upper panels:} Time-evolution of the Mukhanov-Sasaki mass, see Eq.~\eqref{eqn: MS mass}, represented against e-folds, $N$. \textit{Lower panels:} Time-derivative of the Mukhanov-Sasaki mass normalised by $V_{\rm CMB}$, see Eq~\eqref{eqn: m^2 time variation}.
    Each column corresponds to one of the three Hubble-flow models presented in Sec.~\ref{sec: hubble-flow parametrisation}. 
    The black, dashed lines displays quantities computed for the reconstructed potentials in each case, while coloured lines are obtained after UV filtering, see Sec.~\ref{sec: uv filtering}. 
    See Fig.~\ref{fig: V''' power filtered eta(N) models} for the color legend.}
    \label{fig: sqrt(eps1)V''' filtered eta(N) models}
\end{figure}

We display in Fig.~\ref{fig: sqrt(eps1)V''' filtered eta(N) models} the time dependence of the Mukhanov--Sasaki mass and $\sqrt{\epsilon_1} \, V_{\phi\phi\phi}$, see Eq.~\eqref{eqn: m^2 time variation}, for the Hubble-flow models of Sec.~\ref{sec: hubble-flow parametrisation}, both in their original form and after UV filtering. 
Let us discuss the results for the original potentials first. 
For all cases the Mukhanov--Sasaki mass, $\mu^2$, varies across the first transition from SR to non-attractor. 
At the second transition from non-attractor back to attractor, it remains approximately constant for Case $\mathbb{I}$, approaching the late-time plateau from below already during the non-attractor phase.  
On the other hand, for cases $\mathbb{II}$ and $\mathbb{III}$ $\mu^2$ varies during the second transition, approaching from below the late-time value in the final attractor phase (Case $\mathbb{II}$) or from above (Case $\mathbb{III}$).
This behaviour is consistent with the results for $V_{\phi\phi\phi}$ displayed in Fig.~\ref{fig:potentials original eta model} and those for $\sqrt{\epsilon_1}\,V_{\phi\phi\phi}$ displayed in the lower panels in Fig.~\ref{fig: sqrt(eps1)V''' filtered eta(N) models}.
Case $\mathbb{I}$ is the model with smallest $V_{\phi \phi \phi}$ across the second transition, and in this case $\sqrt{\epsilon_1}\,V_{\phi\phi\phi}$ therefore remains small.
Case $\mathbb{I}$ realises WDI, and this explains the label given to this case in Sec.~\ref{sec: hubble-flow parametrisation}.  
On the other hand, the sharp spikes in $V_{\phi\phi\phi}$ observed for cases $\mathbb{II}$ and $\mathbb{III}$ cause the Mukhanov--Sasaki mass to vary in time, see Eq.~\eqref{eqn: m^2 time variation}, and lead to the breaking of WDI. 
 
The colored lines in Fig.~\ref{fig: sqrt(eps1)V''' filtered eta(N) models} show the effect that the UV filters introduced in Sec.~\ref{sec: uv filtering} have on the Mukhanov–Sasaki mass and its time variation. 
Filtering high-frequency field-space modes smooths $V_{\phi\phi\phi}$ across the second transition, see Fig.~\ref{fig: filtered potentials eta model}. 
Except for filters with very low $k_{\phi,c}$, the removal of UV power approximately restores WDI for cases $\mathbb{II}$ and $\mathbb{III}$. 
In the bottom panels, one correspondingly observes a gradual flattening of the bump near the second transition, caused in the original potentials by the large $V_{\phi\phi\phi}$.
On the other hand, for Case $\mathbb{I}$, $\mu^2$ begins to slowly vary during the transition from the non-attractor to the final attractor phase as UV power is removed and the duality relation is broken.
The reason for the time-variation of the Mukhanov--Sasaki mass even when $V_{\phi\phi\phi}$ is smoothed out is that the filtering simultaneously makes the potential less flat.
This induces a faster rolling of the field, i.e.\ a larger $\epsilon_1$, and consequently $\sqrt{\epsilon_1} \,V_{\phi\phi\phi}$ is no longer sufficiently suppressed.
Note that for very low $k_{\phi,c}$, the UV filter also starts affecting the transition from SR to non-attractor in all models. 

\begin{figure}
    \centering
    \begin{subfigure}{0.48\textwidth}
        \centering
        \includegraphics[width=\textwidth]{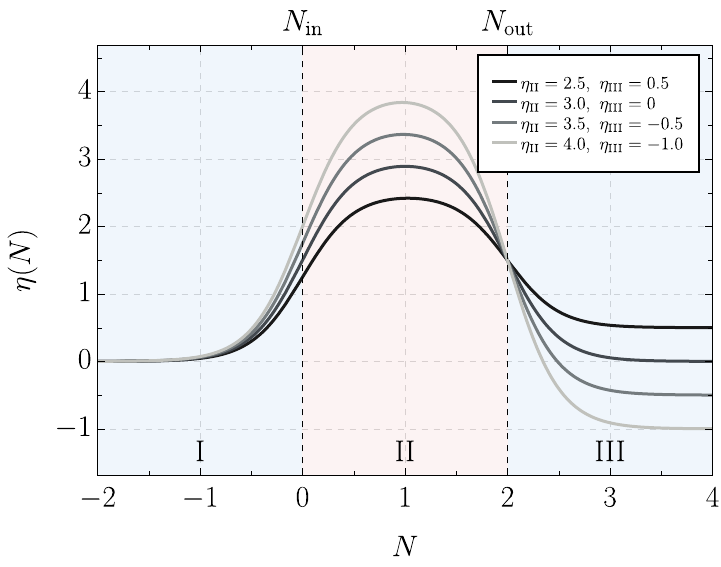}
        \label{fig:eta_sweep}
    \end{subfigure}
    \hfill
    \begin{subfigure}{0.48\textwidth}
        \centering
        \includegraphics[width=\textwidth]{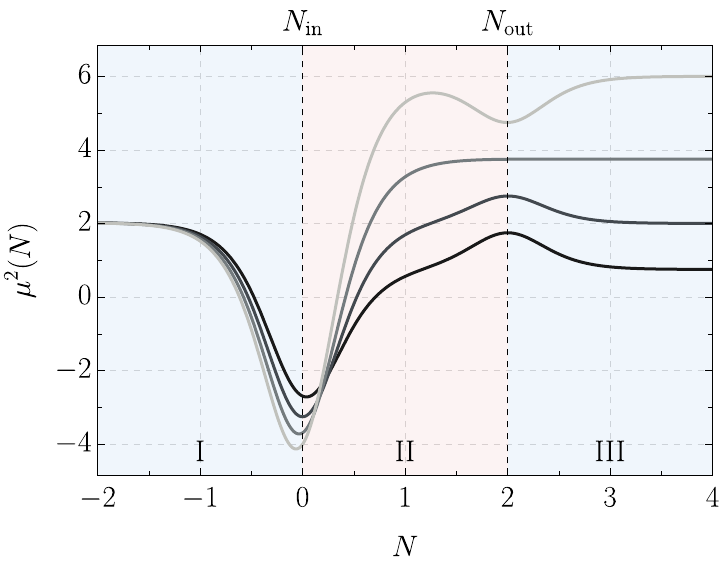}
        \label{fig:plot_ms}
    \end{subfigure}
    \caption{\textit{Left panel:} Time-evolution of $\eta(N)$ for four Hubble-flow models, see Eq.~\eqref{eqn: eta parametrisation}. 
    We set $\eta_{\rm I}=0$, $N_{\rm in}=0$, $\Delta N = 2$ and $\delta N = 0.5$ and vary $\eta_{\rm II}$ and $\eta_{\rm III}$ such that the condition~\eqref{eqn: background transf WDI} is realised. 
    The model with $\eta_{\rm II}=3.5$ and $\eta_{\rm III}=-0.5$ is Case $\mathbb{I}$ of Fig.~\ref{fig: etas original eta model}. 
    \textit{Right panel:} Time-evolution of the Mukhanov--Sasaki mass, see Eq.~\eqref{eqn: MS mass} for the four models represented in the left panel.}
    \label{fig:combined_profiles}
\end{figure}

Equation~\eqref{eqn: eta parametrisation} allows one to choose the values of $\eta$ during the non-attractor and last attractor eras independently.
For Case $\mathbb{I}$ they satisfy Eq.~\eqref{eqn: background transf WDI}, while they do not for cases $\mathbb{II}$ and $\mathbb{III}$, see Table~\ref{tab:USR_cases}. 
One might wonder whether one can construct a Hubble-flow model~\eqref{eqn: eta parametrisation} that respects WDI, such as Case $\mathbb{I}$, simply by choosing the values of $\eta_{\rm II}$ and $\eta_{\rm III}$ such that Eq.~\eqref{eqn: background transf WDI} is satisfied.
However, Eq.~\eqref{eqn: background transf WDI} is not sufficient by itself to realise WDI, and Eq.~\eqref{eqn: m^2 time variation} is the condition leading to WDI in the decoupling limit ($\epsilon_1\to0$).
We demonstrate this in Fig.~\ref{fig:combined_profiles}.
Here we represent four $\eta(N)$ models~\eqref{eqn: eta parametrisation} with all other parameters fixed while varying $\eta_{\rm II}$ and $\eta_{\rm III}$ such that they respect Eq.~\eqref{eqn: background transf WDI}.
The model with $\eta_{\rm II}=3.5$ and $\eta_{\rm III}=-0.5$ is Case $\mathbb{I}$ of Sec.~\ref{sec: hubble-flow parametrisation}. 
While all models respect the condition~\eqref{eqn: background transf WDI}, only Case $\mathbb{I}$ realises WDI at the second transition, as demonstrated by the time-evolution of the Mukhanov--Sasaki mass displayed in the right panel. 
In other words, further tuning of the other Hubble-flow parameters in Eq.~\eqref{eqn: eta parametrisation} is needed in order to ensure that the  Mukhanov--Sasaki mass remains constant across the second transition.

\begin{figure}
    \centering
    \includegraphics[width=1.0\linewidth]{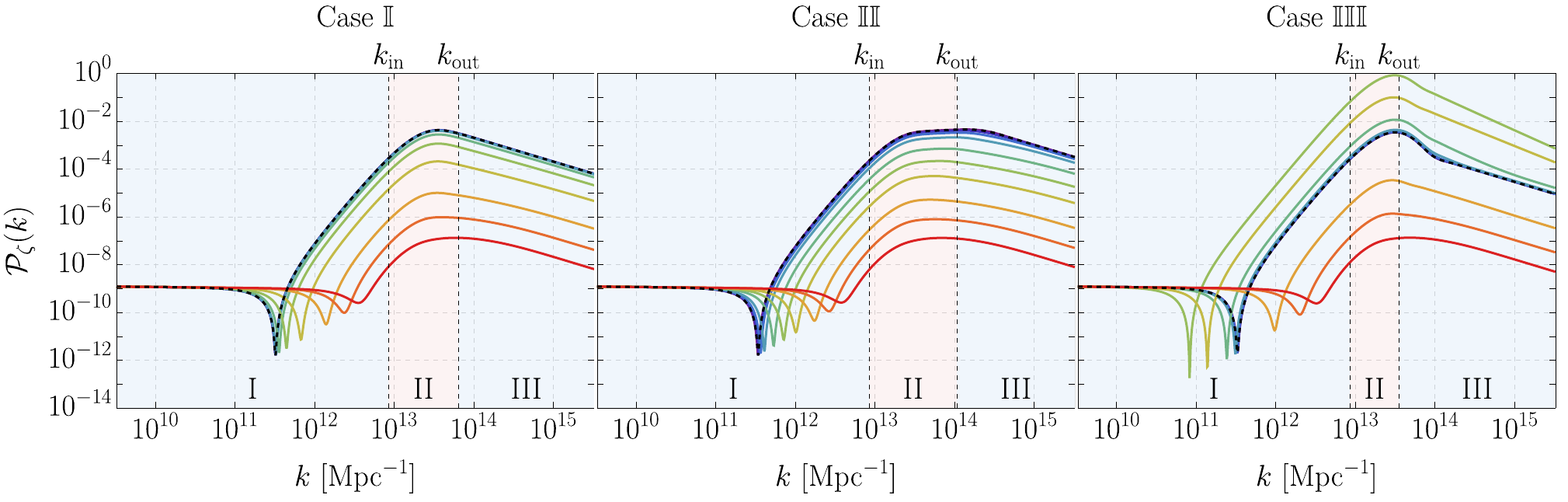}
    \caption{Primordial scalar power spectrum, $\mathcal{P}_\zeta(k)$, as a function of comoving wavenumber $k$, for the Hubble-flow models of Sec.~\ref{sec: hubble-flow parametrisation}. 
    The comoving wavenumber $k$ is normalised such that the CMB horizon-crossing scale matches the CMB pivot scale $0.05\,\mathrm{Mpc}^{-1}$. The plot only shows the region near the peak. 
    The black, dashed lines display quantities computed for the reconstructed potentials in each case, while coloured lines are obtained after UV filtering, see Sec.~\ref{sec: uv filtering}. 
    See Fig.~\ref{fig: V''' power filtered eta(N) models} for the color legend.
    The vertical dashed lines mark $k_{\rm in}=a(N_{\rm in})H(N_{\rm in})$ and $k_{\rm out}=a(N_{\rm out})H(N_{\rm out})$ respectively.}
    \label{fig: p(k) filtered eta models}
\end{figure}

In Fig.~\ref{fig: p(k) filtered eta models} we show numerical results for the primordial scalar power spectrum, $\mathcal{P}_\zeta(k)$.
For the original potential of Case $\mathbb{I}$, the time dependence of the MS mass shown in Fig.~\ref{fig: sqrt(eps1)V''' filtered eta(N) models} shows only one transition during inflation, i.e. the transition from slow roll to non-attractor phase~\cite{Jackson:2023obv}.
The system is no longer almost scale-invariant (as it would be in slow roll). 
In the spectrum of perturbations the appearance of a new energy scale is manifested as a broken-power-law peak, characterised by a single scale, $k_\text{peak}$. 
On the other hand, for cases $\mathbb{II}$ and $\mathbb{III}$ the breaking of WDI indicates that there are now \textit{two} transitions evident in the MS mass, with two new associated energy scales.
The spectra for cases $\mathbb{II}$ and $\mathbb{III}$ have a richer peak structure, characterised by two consecutive breaks of the power-law behavior. 

Once the UV filter is applied to these Hubble-flow models, the overall amplitude of the power spectrum is reduced, excluding some mild filters indicated in green in case $\mathbb{III}$. 
This can be understood by considering that the super-horizon growth of $\zeta$ is determined by the integrated effect of  $\epsilon_2(N)\simeq -2\eta(N)$, see Eq.~\eqref{eqn: zeta super-horizon solution1}. 
The UV filter progressively reduces the amplitude and length of the non-attractor phase, see Fig.~\ref{fig: eta and eps1 filtered eta models}, and $\zeta$ is therefore less enhanced.
Also, we observe that the second breaking of power-law behavior observed for cases $\mathbb{II}$ and $\mathbb{III}$ gradually disappears.
The second feature in the power spectrum associated with the second transition is removed by the UV filter because WDI is approximately restored.

Our results in Figs.~\ref{fig: sqrt(eps1)V''' filtered eta(N) models} and~\ref{fig: p(k) filtered eta models} demonstrate that for cases $\mathbb{II}$ and $\mathbb{III}$ the breaking of WDI and the corresponding appearance of two characteristic scales in the peak of $\mathcal{P}_\zeta$ are due to the sharp spikes in the corresponding potentials.

\subsubsection{The case of analytical potentials}
\label{sec: WDI potential models}

\begin{figure}
    \centering
    \includegraphics[width=1.0\linewidth]{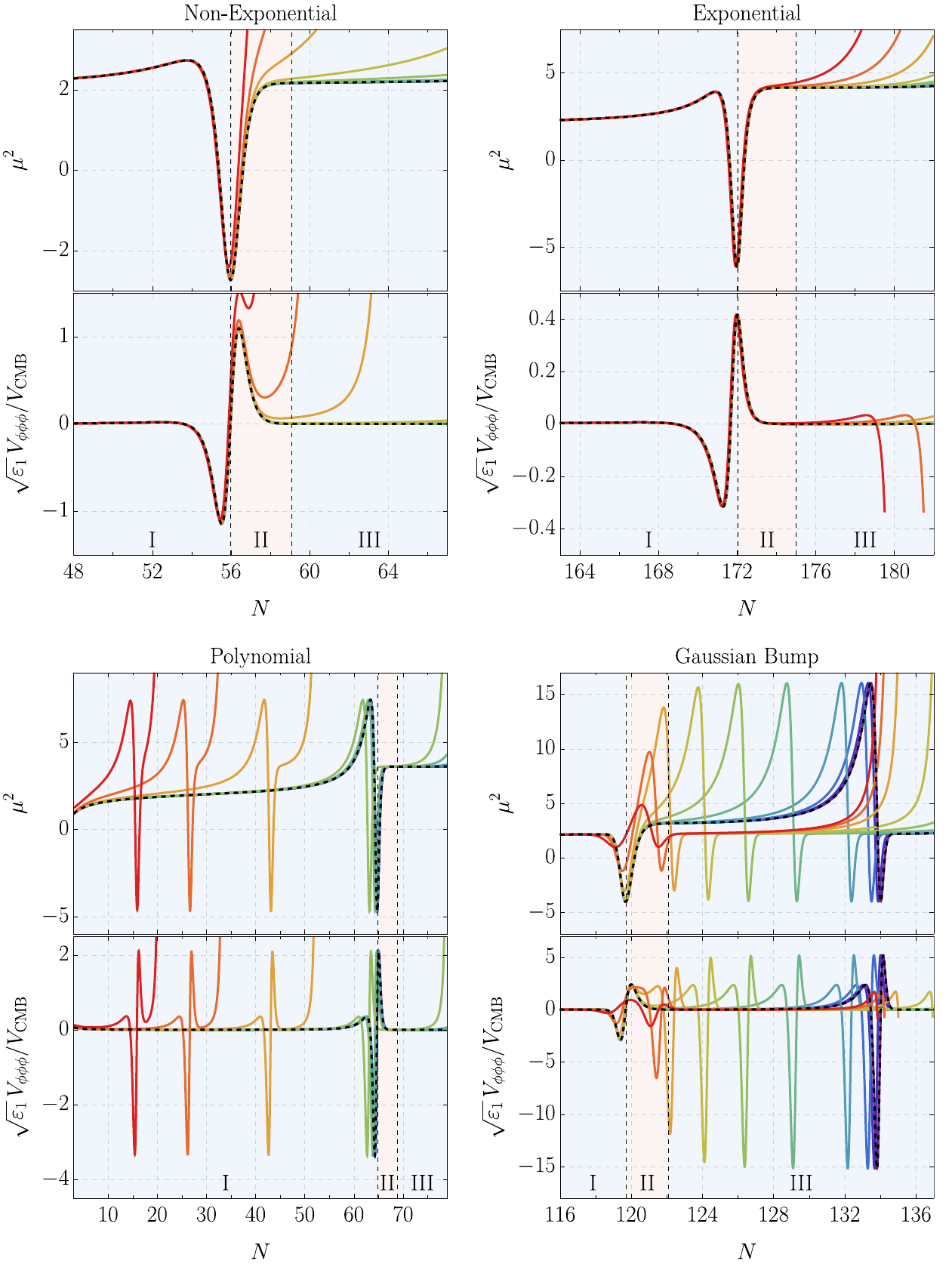}
    \caption{\textit{Upper panels:} Time-evolution of the Mukhanov--Sasaki mass, see Eq.~\eqref{eqn: MS mass}, represented against e-folds, $N$. 
    \textit{Lower panels:} Time-derivative of the Mukhanov--Sasaki mass normalised by $V_\text{CMB}$, see Eq.~\eqref{eqn: m^2 time variation}. 
    Each one of the four quadrants corresponds to one of the analytical models presented in Sec.~\ref{sec: UV Filtering analytical potentials models}. 
    The black, dashed line displays quantities computed for the original potentials, while coloured lines are obtained after UV filtering. 
    See Fig.~\ref{fig: V''' power filtered eta(N) models} for the color legend. The red shaded region corresponds to the non-attractor phase, $\eta \gtrsim 3/2$, of the unfiltered model.}
    \label{fig: m^2 and sqrt(eps1) analitycal models}
\end{figure}

We display in Fig.~\ref{fig: m^2 and sqrt(eps1) analitycal models} the time dependence of the Mukhanov--Sasaki mass and $\sqrt{\epsilon_1}\, V_{\phi\phi\phi}$ for the analytical potentials of Sec.~\ref{sec: UV Filtering analytical potentials models}.
Let us first discuss results obtained for the original potentials. 
The Mukhanov--Sasaki mass does not vary during the second transition from the non-attractor phase back to SR, and all models satisfy WDI between the non-attractor phase and the subsequent attractor era.
This is also demonstrated by the lower panels, where one sees $\sqrt{\epsilon_1} V_{\phi\phi\phi} \to 0$ at the second transition. 
These results are consistent with the characteristics of the analytical potentials, with smooth $V_{\phi\phi\phi}$ at the second transition, see Fig.~\ref{fig: analytical potentials and third derivative}. 
From Fig.~\ref{fig: eta and eps1 filtered analytical potentials} one can see that the values of $\eta$ in these two phases are $\{\eta_{\rm NA} =3,\, \eta_{\rm A} = 0\}$ for the non-exponential model, $\{\eta_{\rm NA} =3.5,\, \eta_{\rm A} = -0.5\}$ for the exponential model, $\{\eta_{\rm NA} =3.45,\, \eta_{\rm A} = -0.45\}$ for the polynomial model. 
Note all these values satisfy the condition in Eq.~\eqref{eqn: background transf WDI}. 
The profile for the Gaussian bump model is more complex, as it exhibits more than one transition, see the two dips in the Mukhanov-–Sasaki mass. 
This behaviour reflects the fact that the field slows down as it approaches the center of the Gaussian bump, accelerates again as it rolls down the bump, before slowing down once more in order to rejoin to the baseline slow-roll attractor solution.
During the non-attractor phase, the model exhibits an approximate WDI, with the $\eta$ evolving from $\eta_{\rm NA} = 3.25$ to $\eta_{\rm A} \simeq -0.35$. 
Note that these values do not satisfy precisely Eq.~\eqref{eqn: background transf WDI}. 

By inspecting the colored lines in Fig.~\ref{fig: m^2 and sqrt(eps1) analitycal models}, one observes that when $k_{\phi,c}$ is large the UV filters leave the Mukhanov--Sasaki mass largely unaffected. 
When more UV power is removed, the filter smooths out the inflection point in the potential.
The system never reaches and settles in the last attractor regime and inflation ends earlier. 
These results are consistent with the behaviour of the SR parameters observed in Fig.~\ref{fig: eta and eps1 filtered analytical potentials}. 

\begin{figure}
    \centering
    \includegraphics[width=1.0\linewidth]{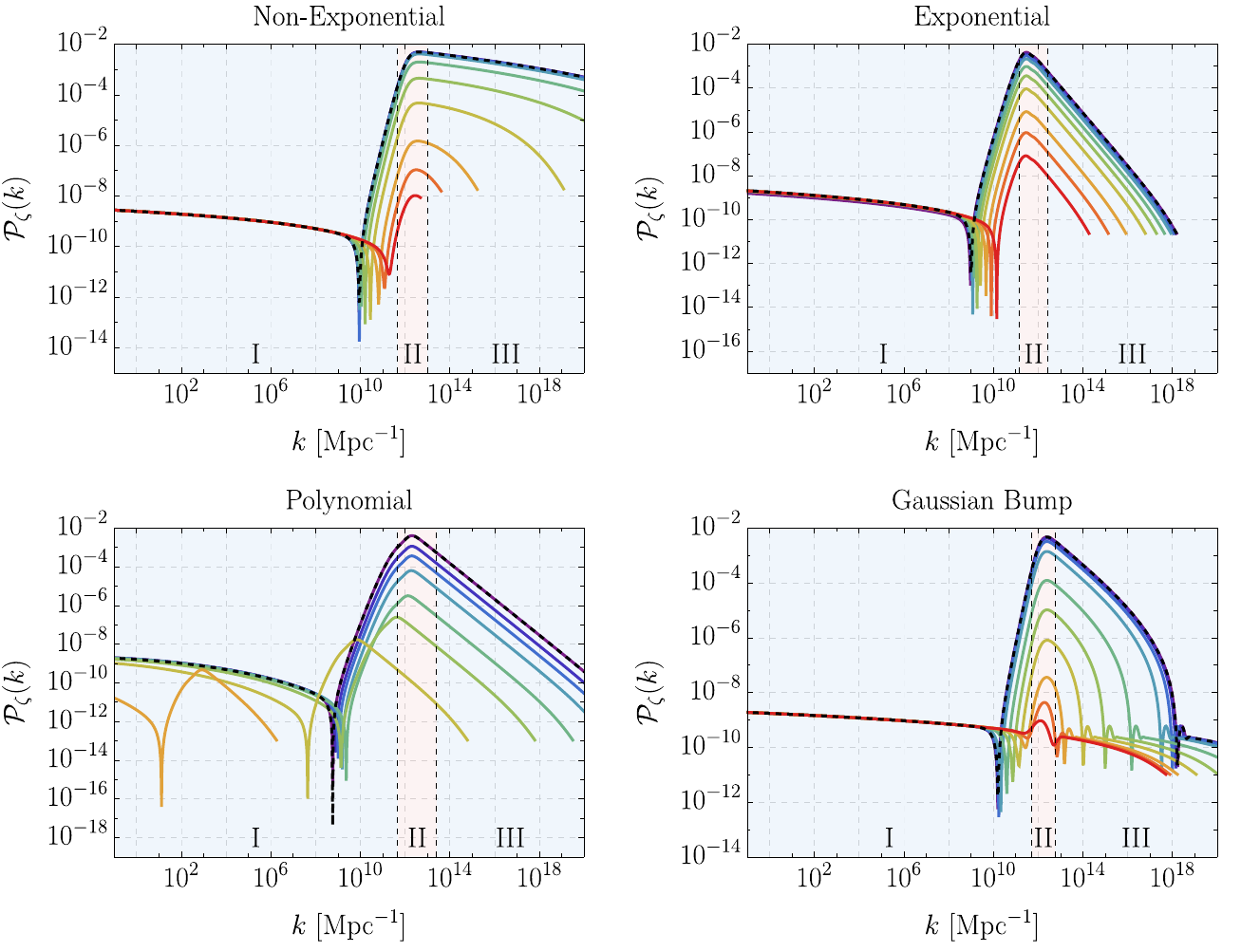}
     \caption{Primordial scalar power spectrum, $\mathcal{P}_\zeta(k)$, computed numerically for the analytical potentials of Sec.~\ref{sec: UV Filtering analytical potentials models}.
     The comoving wavenumber $k$ is normalised such that the CMB horizon-crossing scale matches the CMB pivot scale $0.05\, \mathrm{Mpc}^{-1}$. 
     The back, dashed lines represent results computed from the original potential in each case, while colored lines display those obtained after UV filtering.
     See Fig.~\ref{fig: V''' power filtered eta(N) models} for the color legend. The vertical, dashed lines mark $k=aH$ at the beginning and end of the non-attractor phase of the unfilered model.}
     \label{fig: P(k) analytical potentials}
\end{figure}

In Fig.~\ref{fig: P(k) analytical potentials}, we present numerical results for the primordial scalar power spectrum, $\mathcal{P}_\zeta(k)$. 
The spectra obtained without UV filtering display a broken-power-law peak, consistent with the fact that the Mukhanov--Sasaki mass only displays one sudden transition~\cite{Jackson:2023obv}. 
The only exception is the Gaussian bump model, which after the principal peak shows a second feature, much reduced in amplitude. 
The richer structure of the spectrum in this case is due to the presence of a second transition after the inflaton has joined an intermediate attractor phase after the non-attractor transient era.
 
As anticipated, the maximum value of $\eta(N)$ during the non-attractor phase is suppressed, leading to an overall reduction in $\mathcal{P}_{\zeta}(k)$. Apart from the polynomial potential, where the dynamics are shifted to earlier times, mild filtering preserves the shape of the spectrum. In contrast to Hubble-flow-derived parametrisations, no secondary transition feature is present.

Our results in Figs.~\ref{fig: m^2 and sqrt(eps1) analitycal models} and~\ref{fig: P(k) analytical potentials} show that explicit microphysical models of non-attractor inflation typically realise WDI and their spectra display one characteristic scale only. 
The progressive removal of UV power does not modify these characteristics, unless the potential is so flattened by the filter that inflation ends before the inflaton can relax to the final SR attractor solution. 
While $\eta(N)$ models~\eqref{eqn: eta parametrisation} can lead to breaking of WDI, this is typically not observed in analytical models due to the smooth nature of these potentials. 

\section{Discussion}
\label{sec: discussion}

In this work, we have critically examined the modeling of transient non-attractor inflation through an analytical Hubble-flow parametrisation, see Eq.~\eqref{eqn: eta parametrisation}. 
We find that the underlying potentials may include specific, highly-localised spikes in its higher-derivatives that are absent in typical potential models.

We consider the three Hubble-flow models studied in Refs.~\cite{Caravano:2024moy, Caravano:2025diq,Caravano:2026hca}, labeled cases $\mathbb{I}$, $\mathbb{II}$ and $\mathbb{III}$ respectively.
While the time-evolution of $\eta(N)$ looks similar for the three models considered, see Fig.~\ref{fig: etas original eta model}, Case $\mathbb{I}$ is special in that the Hubble-flow parameters are chosen in order to satisfy Wands duality.
In Sec.~\ref{sec: reconstruction of the inflationary potential} we reconstructed the underlying inflationary potentials that lead to the specified dynamics in each case, see Fig.~\ref{fig:potentials original eta model}. 
While the potentials, $V(\phi)$, appear smooth in each case, cases $\mathbb{II}$ and $\mathbb{III}$ display sharp features in $V_{\phi\phi\phi}$ at the time of the transition from the non-attractor phase back to the last attractor era. 
In case $\mathbb{I}$, $V_{\phi\phi\phi}$ remains small compared to the spike seen in cases $\mathbb{II}$ and $\mathbb{III}$, but there is nonetheless an abrupt jump in its value at the transition. 
Such UV features are due to the imposed background dynamics encoded in $\eta(N)$, and are absent in typical analytical potentials proposed in the literature.
This can be seen in Figs.~\ref{fig: V''' power filtered eta(N) models} and~\ref{fig: dft pwer spectrum analytical potentials}, where we represent the power spectrum of the discrete Fourier transform of $V_{\phi\phi\phi}$ computed for the reconstructed potentials and for four analytical $V(\phi)$ models discussed in Ref.~\cite{Cole:2023wyx}.  
While for typical analytical potentials and for case $\mathbb{I}$ the spectrum of $V_{\phi\phi\phi}$ is red tilted in the UV, the reconstructed potentials for cases $\mathbb{II}$ and $\mathbb{III}$ display an excess of UV power.  
These results show that Hubble-flow parametrisations are far from being model-independent. 
On the contrary, the imposed background dynamics (and in particular the transition from the non-attractor phase to the final attractor era) are realised by atypical inflationary potentials (compared to analytic $V(\phi)$ models) with spurious UV  artefacts\footnote{We refer to these features as artefacts in the sense that they arise as a direct consequence of the Hubble-flow parameterisation. We do not mean to imply that they are a result of any analytical or numerical error in the process.} in higher-order derivatives.

By UV-filtering $V_{\phi \phi \phi}$, we are able to assess the impact of sharp features on the primordial power spectrum, and distinguish its underlying shape from parametrisation artefacts.
Case $\mathbb{I}$ satisfies Wands duality by construction, and its primordial power spectrum, shown in Fig.~\ref{fig: p(k) filtered eta models}, is well described by a broken power-law.
Once applied to the potential in case $\mathbb{I}$, the UV filter induces a mild breaking of duality invariance at the second transition. 
For cases $\mathbb{II}$ and $\mathbb{III}$ spurious UV features in $V_{\phi\phi\phi}$ lead to violation of Wands duality invariance between the non-attractor and subsequent attractor phases.
This is manifested as a time-varying mass for the Mukhanov--Sasaki variable at the time of the second transition, in contrast to case $\mathbb{I}$; see the black lines in Fig.~\ref{fig: sqrt(eps1)V''' filtered eta(N) models}. 
As a result, the peak in the scalar power spectrum $\mathcal{P}_\zeta(k)$ displays a rich structure, with two consecutive breakings of power-law behavior, see the black lines in Fig.~\ref{fig: p(k) filtered eta models}. 
By applying the filter to the reconstructed potentials for cases $\mathbb{II}$ and $\mathbb{III}$, UV features in $V_{\phi\phi\phi}$ are suppressed, see Fig.~\ref{fig: filtered potentials eta model}. 
The progressive cut of UV power in $V_{\phi\phi\phi}$ smooths the time evolution of the Mukhanov--Sasaki mass, thereby approximately restoring Wands duality invariance (see Fig.~\ref{fig: sqrt(eps1)V''' filtered eta(N) models}) and smooths the breakings of power-law behavior in $\mathcal{P}_\zeta(k)$ (see Fig.~\ref{fig: p(k) filtered eta models}). 

In order to compare these Hubble-flow models against realisations of non-attractor inflation from explicit potentials, we consider the microphysical models discussed in Ref.~\cite{Cole:2023wyx} and apply the UV filter to them. 
We show that analytical potentials tend to display Wands duality invariance when they relax from the non-attractor phase back to an attractor trajectory, see Fig.~\ref{fig: m^2 and sqrt(eps1) analitycal models}. 
For these analytical models, the spectral shape of $\mathcal{P}_\zeta(k)$ is typically more stable under the action of the UV filter, see Fig.~\ref{fig: P(k) analytical potentials}. 
On the other hand, both for $\eta(N)$ and $V(\phi)$ models, the amplitude and position of the peak in $\mathcal{P}_\zeta(k)$ are quite sensitive to the action of the filter. 
This reflects the extreme sensitivity of background dynamics in models with non-attractor behavior; even mild suppression of high-frequency modes in the potential can significantly alter the duration of the non-attractor phase, often causing a early end to inflation. 
The SR parameters are shown in Figs.~\ref{fig: eta and eps1 filtered eta models} and~\ref{fig: eta and eps1 filtered analytical potentials} for the UV-filtered Hubble-flow potentials and filtered analytical potentials respectively. 
This sensitivity highlights the fine-tuning required to sustain a sufficiently long non-attractor phase (necessary to amplify the primordial power spectrum enough for PBH formation), regardless of whether the potential is specified analytically or reconstructed from a Hubble-flow parametrisation.

Our work shows that apparently simple Hubble-flow parametrisations may not be suitable for modeling inflation with a transient non-attractor phase.  
In particular, the background dynamics imposed at the transition to the final attractor phase requires underlying potentials with sharp, localised spikes in higher-order derivatives. 
Previous work has shown in other contexts that apparently generic predictions from Hubble-flow parameterisations may result from highly model-specific choice of potentials~\cite{Vennin:2014xta}. 
While in this work we have explored the effect of hidden UV features on linear perturbations, we expect higher-order correlators also to be affected~\cite{Cai:2018dkf, Kristiano:2024vst}.  
This is because higher derivatives of the inflationary potential control the size of non-linear interactions.  
Hyperbolic tangent profiles for $\eta(N)$ (such as the one examined in this work) have been widely adopted in computing non-Gaussianities~\cite{Taoso:2021uvl, Ballesteros:2024pbe, Ballesteros:2024pwn} and the power spectrum beyond linear order, both by means of non-perturbative lattice simulations~\cite{Caravano:2024moy, Caravano:2025diq, Caravano:2026hca} and in a perturbative fashion within a loop expansion~\cite{Franciolini:2023agm, Ballesteros:2024zdp}. 
Consequently, we should be wary of drawing conclusions about nonlinear effects in ultra-slow-roll models if the predictions are sensitive to UV features in the underlying potential, $V(\phi)$, introduced by the Hubble-flow parameterisation.

\acknowledgments
We would like to thank V. Vennin for careful reading of the manuscript and very useful comments.  
The authors are grateful to A.~Caravano, D.~Cruces, G.~Franciolini, D. Mulryne and S.~Renaux-Petel for interesting and useful discussions related to this work. 
GB is supported by STFC grant number UKRI1796.
LI is supported by STFC grant number ST/X000931/1. 
KK is  supported by STFC grant number ST/B001175/1.
For the purpose of open access, the authors have applied a Creative Commons Attribution (CC-BY) licence to any Author Accepted Manuscript version arising from this work.
Supporting research data are available on reasonable request. 

\bibliography{refs} 

@article{Jackson:2023obv,
    author = "Jackson, Joseph H. P. and Assadullahi, Hooshyar and Gow, Andrew D. and Koyama, Kazuya and Vennin, Vincent and Wands, David",
    title = "{The separate-universe approach and sudden transitions during inflation}",
    eprint = "2311.03281",
    archivePrefix = "arXiv",
    primaryClass = "astro-ph.CO",
    doi = "10.1088/1475-7516/2024/05/053",
    journal = "JCAP",
    volume = "05",
    pages = "053",
    year = "2024"
}

@article{Iacconi:2021ltm,
    author = "Iacconi, Laura and Assadullahi, Hooshyar and Fasiello, Matteo and Wands, David",
    title = "{Revisiting small-scale fluctuations in {\ensuremath{\alpha}}-attractor models of inflation}",
    eprint = "2112.05092",
    archivePrefix = "arXiv",
    primaryClass = "astro-ph.CO",
    doi = "10.1088/1475-7516/2022/06/007",
    journal = "JCAP",
    volume = "06",
    number = "06",
    pages = "007",
    year = "2022"
}

@article{Leach:2001zf,
    author = "Leach, Samuel M and Sasaki, Misao and Wands, David and Liddle, Andrew R",
    title = "{Enhancement of superhorizon scale inflationary curvature perturbations}",
    eprint = "astro-ph/0101406",
    archivePrefix = "arXiv",
    doi = "10.1103/PhysRevD.64.023512",
    journal = "Phys. Rev. D",
    volume = "64",
    pages = "023512",
    year = "2001"
}

@article{Ragavendra:2024yfp,
    author = "Ragavendra, H. V. and Sarkar, Anjan Kumar and Sethi, Shiv K.",
    title = "{Constraining ultra slow roll inflation using cosmological datasets}",
    eprint = "2404.00933",
    archivePrefix = "arXiv",
    primaryClass = "astro-ph.CO",
    doi = "10.1088/1475-7516/2024/07/088",
    journal = "JCAP",
    volume = "07",
    pages = "088",
    year = "2024"
}

@article{Carr:2026hot,
    author = {Carr, Bernard and Iovino, Antonio J. and Perna, Gabriele and Vaskonen, Ville and Veerm{\"a}e, Hardi},
    title = "{Primordial black holes: constraints, potential evidence and prospects}",
    eprint = "2601.06024",
    archivePrefix = "arXiv",
    primaryClass = "astro-ph.CO",
    doi = "10.1007/s40766-026-00080-z",
    month = "1",
    year = "2026"
}

@article{Caravano:2024moy,
    author = "Caravano, Angelo and Franciolini, Gabriele and Renaux-Petel, S{\'e}bastien",
    title = "{Ultraslow-roll inflation on the lattice: Backreaction and nonlinear effects}",
    eprint = "2410.23942",
    archivePrefix = "arXiv",
    primaryClass = "astro-ph.CO",
    reportNumber = "CERN-TH-2024-181",
    doi = "10.1103/PhysRevD.111.063518",
    journal = "Phys. Rev. D",
    volume = "111",
    number = "6",
    pages = "063518",
    year = "2025"
}

@article{Cole:2023wyx,
    author = "Cole, Philippa S. and Gow, Andrew D. and Byrnes, Christian T. and Patil, Subodh P.",
    title = "{Primordial black holes from single-field inflation: a fine-tuning audit}",
    eprint = "2304.01997",
    archivePrefix = "arXiv",
    primaryClass = "astro-ph.CO",
    doi = "10.1088/1475-7516/2023/08/031",
    journal = "JCAP",
    volume = "08",
    pages = "031",
    year = "2023"
}

@article{Caravano:2025diq,
    author = "Caravano, Angelo and Franciolini, Gabriele and Renaux-Petel, S{\'e}bastien",
    title = "{Ultraslow-roll inflation on the lattice. II. Nonperturbative curvature perturbation}",
    eprint = "2506.11795",
    archivePrefix = "arXiv",
    primaryClass = "astro-ph.CO",
    reportNumber = "CERN-TH-2025-116",
    doi = "10.1103/39qd-gdfm",
    journal = "Phys. Rev. D",
    volume = "112",
    number = "8",
    pages = "083508",
    year = "2025"
}

@article{Caravano:2026hca,
    author = "Caravano, Angelo and Franciolini, Gabriele and Renaux-Petel, S{\'e}bastien",
    title = "{Lattice simulations of scalar-induced gravitational waves from inflation}",
    eprint = "2604.03628",
    archivePrefix = "arXiv",
    primaryClass = "astro-ph.CO",
    month = "4",
    year = "2026"
}

@article{Wands:1998yp,
    author = "Wands, David",
    title = "{Duality invariance of cosmological perturbation spectra}",
    eprint = "gr-qc/9809062",
    archivePrefix = "arXiv",
    reportNumber = "PU-RCG-98-15",
    doi = "10.1103/PhysRevD.60.023507",
    journal = "Phys. Rev. D",
    volume = "60",
    pages = "023507",
    year = "1999"
}

@article{Ballesteros:2024zdp,
    author = "Ballesteros, Guillermo and Gamb{\'\i}n Egea, Jes{\'u}s",
    title = "{One-loop power spectrum in ultra slow-roll inflation and implications for primordial black hole dark matter}",
    eprint = "2404.07196",
    archivePrefix = "arXiv",
    primaryClass = "astro-ph.CO",
    doi = "10.1088/1475-7516/2024/07/052",
    journal = "JCAP",
    volume = "07",
    pages = "052",
    year = "2024"
}

@article{Atal:2018neu,
    author = "Atal, Vicente and Germani, Cristiano",
    title = "{The role of non-gaussianities in Primordial Black Hole formation}",
    eprint = "1811.07857",
    archivePrefix = "arXiv",
    primaryClass = "astro-ph.CO",
    reportNumber = "ICCUB-18-022",
    doi = "10.1016/j.dark.2019.100275",
    journal = "Phys. Dark Univ.",
    volume = "24",
    pages = "100275",
    year = "2019"
}

@article{Bezrukov:2014bra,
    author = "Bezrukov, Fedor and Shaposhnikov, Mikhail",
    title = "{Higgs inflation at the critical point}",
    eprint = "1403.6078",
    archivePrefix = "arXiv",
    primaryClass = "hep-ph",
    reportNumber = "CERN-PH-TH-2014-082",
    doi = "10.1016/j.physletb.2014.05.074",
    journal = "Phys. Lett. B",
    volume = "734",
    pages = "249--254",
    year = "2014"
}

@article{Garcia-Bellido:2017mdw,
    author = "Garcia-Bellido, Juan and Ruiz Morales, Ester",
    title = "{Primordial black holes from single field models of inflation}",
    eprint = "1702.03901",
    archivePrefix = "arXiv",
    primaryClass = "astro-ph.CO",
    reportNumber = "IFT-UAM-CSIC-17-007, CERN-TH-2017-196",
    doi = "10.1016/j.dark.2017.09.007",
    journal = "Phys. Dark Univ.",
    volume = "18",
    pages = "47--54",
    year = "2017"
}

@article{Germani:2017bcs,
    author = "Germani, Cristiano and Prokopec, Tomislav",
    title = "{On primordial black holes from an inflection point}",
    eprint = "1706.04226",
    archivePrefix = "arXiv",
    primaryClass = "astro-ph.CO",
    reportNumber = "ICCUB-17-012",
    doi = "10.1016/j.dark.2017.09.001",
    journal = "Phys. Dark Univ.",
    volume = "18",
    pages = "6--10",
    year = "2017"
}

@article{Cicoli:2018asa,
    author = "Cicoli, Michele and Diaz, Victor A. and Pedro, Francisco G.",
    title = "{Primordial Black Holes from String Inflation}",
    eprint = "1803.02837",
    archivePrefix = "arXiv",
    primaryClass = "hep-th",
    doi = "10.1088/1475-7516/2018/06/034",
    journal = "JCAP",
    volume = "06",
    pages = "034",
    year = "2018"
}

@article{Hertzberg:2017dkh,
    author = "Hertzberg, Mark P. and Yamada, Masaki",
    title = "{Primordial Black Holes from Polynomial Potentials in Single Field Inflation}",
    eprint = "1712.09750",
    archivePrefix = "arXiv",
    primaryClass = "astro-ph.CO",
    doi = "10.1103/PhysRevD.97.083509",
    journal = "Phys. Rev. D",
    volume = "97",
    number = "8",
    pages = "083509",
    year = "2018"
}

@article{Mishra:2019pzq,
    author = "Mishra, Swagat S. and Sahni, Varun",
    title = "{Primordial Black Holes from a tiny bump/dip in the Inflaton potential}",
    eprint = "1911.00057",
    archivePrefix = "arXiv",
    primaryClass = "gr-qc",
    doi = "10.1088/1475-7516/2020/04/007",
    journal = "JCAP",
    volume = "04",
    pages = "007",
    year = "2020"
}

@article{Kristiano:2022maq,
    author = "Kristiano, Jason and Yokoyama, Jun'ichi",
    title = "{Constraining Primordial Black Hole Formation from Single-Field Inflation}",
    eprint = "2211.03395",
    archivePrefix = "arXiv",
    primaryClass = "hep-th",
    reportNumber = "RESCEU-20/22",
    doi = "10.1103/PhysRevLett.132.221003",
    journal = "Phys. Rev. Lett.",
    volume = "132",
    number = "22",
    pages = "221003",
    year = "2024"
}

@article{Firouzjahi:2023aum,
    author = "Firouzjahi, Hassan",
    title = "{One-loop corrections in power spectrum in single field inflation}",
    eprint = "2303.12025",
    archivePrefix = "arXiv",
    primaryClass = "astro-ph.CO",
    doi = "10.1088/1475-7516/2023/10/006",
    journal = "JCAP",
    volume = "10",
    pages = "006",
    year = "2023"
}

@article{Firouzjahi:2023bkt,
    author = "Firouzjahi, Hassan",
    title = "{Revisiting loop corrections in single field ultraslow-roll inflation}",
    eprint = "2311.04080",
    archivePrefix = "arXiv",
    primaryClass = "astro-ph.CO",
    doi = "10.1103/PhysRevD.109.043514",
    journal = "Phys. Rev. D",
    volume = "109",
    number = "4",
    pages = "043514",
    year = "2024"
}

@article{Riotto:2023hoz,
    author = "Riotto, Antonio",
    title = "{The Primordial Black Hole Formation from Single-Field Inflation is Not Ruled Out}",
    eprint = "2301.00599",
    archivePrefix = "arXiv",
    primaryClass = "astro-ph.CO",
    month = "1",
    year = "2023"
}

@article{Kristiano:2023scm,
    author = "Kristiano, Jason and Yokoyama, Jun'ichi",
    title = "{Note on the bispectrum and one-loop corrections in single-field inflation with primordial black hole formation}",
    eprint = "2303.00341",
    archivePrefix = "arXiv",
    primaryClass = "hep-th",
    reportNumber = "RESCEU-3/23",
    doi = "10.1103/PhysRevD.109.103541",
    journal = "Phys. Rev. D",
    volume = "109",
    number = "10",
    pages = "103541",
    year = "2024"
}

@article{Firouzjahi:2023ahg,
    author = "Firouzjahi, Hassan and Riotto, Antonio",
    title = "{Primordial Black Holes and loops in single-field inflation}",
    eprint = "2304.07801",
    archivePrefix = "arXiv",
    primaryClass = "astro-ph.CO",
    doi = "10.1088/1475-7516/2024/02/021",
    journal = "JCAP",
    volume = "02",
    pages = "021",
    year = "2024"
}

@article{Byrnes:2018txb,
    author = "Byrnes, Christian T. and Cole, Philippa S. and Patil, Subodh P.",
    title = "{Steepest growth of the power spectrum and primordial black holes}",
    eprint = "1811.11158",
    archivePrefix = "arXiv",
    primaryClass = "astro-ph.CO",
    doi = "10.1088/1475-7516/2019/06/028",
    journal = "JCAP",
    volume = "06",
    pages = "028",
    year = "2019"
}

@article{Taoso:2021uvl,
    author = "Taoso, Marco and Urbano, Alfredo",
    title = "{Non-gaussianities for primordial black hole formation}",
    eprint = "2102.03610",
    archivePrefix = "arXiv",
    primaryClass = "astro-ph.CO",
    doi = "10.1088/1475-7516/2021/08/016",
    journal = "JCAP",
    volume = "08",
    pages = "016",
    year = "2021"
}

@article{Franciolini:2022pav,
    author = "Franciolini, Gabriele and Urbano, Alfredo",
    title = "{Primordial black hole dark matter from inflation: The reverse engineering approach}",
    eprint = "2207.10056",
    archivePrefix = "arXiv",
    primaryClass = "astro-ph.CO",
    doi = "10.1103/PhysRevD.106.123519",
    journal = "Phys. Rev. D",
    volume = "106",
    number = "12",
    pages = "123519",
    year = "2022"
}

@article{Franciolini:2023agm,
    author = "Franciolini, Gabriele and Iovino, Junior., Antonio and Taoso, Marco and Urbano, Alfredo",
    title = "{Perturbativity in the presence of ultraslow-roll dynamics}",
    eprint = "2305.03491",
    archivePrefix = "arXiv",
    primaryClass = "astro-ph.CO",
    doi = "10.1103/PhysRevD.109.123550",
    journal = "Phys. Rev. D",
    volume = "109",
    number = "12",
    pages = "123550",
    year = "2024"
}

@article{Cai:2018dkf,
    author = "Cai, Yi-Fu and Chen, Xingang and Namjoo, Mohammad Hossein and Sasaki, Misao and Wang, Dong-Gang and Wang, Ziwei",
    title = "{Revisiting non-Gaussianity from non-attractor inflation models}",
    eprint = "1712.09998",
    archivePrefix = "arXiv",
    primaryClass = "astro-ph.CO",
    reportNumber = "MIT-CTP-4974, YITP-17-133",
    doi = "10.1088/1475-7516/2018/05/012",
    journal = "JCAP",
    volume = "05",
    pages = "012",
    year = "2018"
}

@article{Kristiano:2024vst,
    author = "Kristiano, Jason and Yokoyama, Jun'ichi",
    title = "{Comparing sharp and smooth transitions of the second slow-roll parameter in single-field inflation}",
    eprint = "2405.12145",
    archivePrefix = "arXiv",
    primaryClass = "astro-ph.CO",
    reportNumber = "RESCEU-8/24",
    doi = "10.1088/1475-7516/2024/10/036",
    journal = "JCAP",
    volume = "10",
    pages = "036",
    year = "2024"
}

@article{Motohashi:2023syh,
    author = "Motohashi, Hayato and Tada, Yuichiro",
    title = "{Squeezed bispectrum and one-loop corrections in transient constant-roll inflation}",
    eprint = "2303.16035",
    archivePrefix = "arXiv",
    primaryClass = "astro-ph.CO",
    doi = "10.1088/1475-7516/2023/08/069",
    journal = "JCAP",
    volume = "08",
    pages = "069",
    year = "2023"
}

@article{Ballesteros:2024pbe,
    author = "Ballesteros, Guillermo and Gamb{\'\i}n Egea, Jes{\'u}s and Konstandin, Thomas and P{\'e}rez Rodr{\'\i}guez, Alejandro and Pierre, Mathias and Rey, Juli{\'a}n",
    title = "{Intrinsic non-Gaussianity of ultra slow-roll inflation}",
    eprint = "2412.14106",
    archivePrefix = "arXiv",
    primaryClass = "astro-ph.CO",
    reportNumber = "IFT UAM-CSIC 24-183, DESY-24-205",
    doi = "10.1088/1475-7516/2026/01/012",
    journal = "JCAP",
    volume = "01",
    pages = "012",
    year = "2026"
}

@article{Ballesteros:2024pwn,
    author = "Ballesteros, Guillermo and Konstandin, Thomas and P{\'e}rez Rodr{\'\i}guez, Alejandro and Pierre, Mathias and Rey, Juli{\'a}n",
    title = "{Non-Gaussian tails without stochastic inflation}",
    eprint = "2406.02417",
    archivePrefix = "arXiv",
    primaryClass = "astro-ph.CO",
    reportNumber = "IFT-UAM/CSIC-24-75, DESY-24-071",
    doi = "10.1088/1475-7516/2024/11/013",
    journal = "JCAP",
    volume = "11",
    pages = "013",
    year = "2024"
}

@article{Ballesteros:2020sre,
    author = "Ballesteros, Guillermo and Rey, Juli{\'a}n and Taoso, Marco and Urbano, Alfredo",
    title = "{Stochastic inflationary dynamics beyond slow-roll and consequences for primordial black hole formation}",
    eprint = "2006.14597",
    archivePrefix = "arXiv",
    primaryClass = "astro-ph.CO",
    doi = "10.1088/1475-7516/2020/08/043",
    journal = "JCAP",
    volume = "08",
    pages = "043",
    year = "2020"
}

@article{Ragavendra:2020sop,
    author = "Ragavendra, H. V. and Saha, Pankaj and Sriramkumar, L. and Silk, Joseph",
    title = "{Primordial black holes and secondary gravitational waves from ultraslow roll and punctuated inflation}",
    eprint = "2008.12202",
    archivePrefix = "arXiv",
    primaryClass = "astro-ph.CO",
    doi = "10.1103/PhysRevD.103.083510",
    journal = "Phys. Rev. D",
    volume = "103",
    number = "8",
    pages = "083510",
    year = "2021"
}

@article{Yokoyama:1998pt,
    author = "Yokoyama, Jun'ichi",
    title = "{Chaotic new inflation and formation of primordial black holes}",
    eprint = "astro-ph/9802357",
    archivePrefix = "arXiv",
    reportNumber = "YITP-98-10, SU-ITP-98-04",
    doi = "10.1103/PhysRevD.58.083510",
    journal = "Phys. Rev. D",
    volume = "58",
    pages = "083510",
    year = "1998"
}

@article{Starobinsky:1979ty,
    author = "Starobinsky, Alexei A.",
    editor = "Khalatnikov, I. M. and Mineev, V. P.",
    title = "{Spectrum of relict gravitational radiation and the early state of the universe}",
    journal = "JETP Lett.",
    volume = "30",
    pages = "682--685",
    year = "1979"
}

@article{Starobinsky:1980te,
    author = "Starobinsky, Alexei A.",
    editor = "Khalatnikov, I. M. and Mineev, V. P.",
    title = "{A New Type of Isotropic Cosmological Models Without Singularity}",
    doi = "10.1016/0370-2693(80)90670-X",
    journal = "Phys. Lett. B",
    volume = "91",
    pages = "99--102",
    year = "1980"
}

@article{Guth:1980zm,
    author = "Guth, Alan H.",
    editor = "Fang, Li-Zhi and Ruffini, R.",
    title = "{The Inflationary Universe: A Possible Solution to the Horizon and Flatness Problems}",
    reportNumber = "SLAC-PUB-2576",
    doi = "10.1103/PhysRevD.23.347",
    journal = "Phys. Rev. D",
    volume = "23",
    pages = "347--356",
    year = "1981"
}

@article{Linde:1981mu,
    author = "Linde, Andrei D.",
    editor = "Fang, Li-Zhi and Ruffini, R.",
    title = "{A New Inflationary Universe Scenario: A Possible Solution of the Horizon, Flatness, Homogeneity, Isotropy and Primordial Monopole Problems}",
    reportNumber = "LEBEDEV-81-229",
    doi = "10.1016/0370-2693(82)91219-9",
    journal = "Phys. Lett. B",
    volume = "108",
    pages = "389--393",
    year = "1982"
}

@article{Albrecht:1982wi,
    author = "Albrecht, Andreas and Steinhardt, Paul J.",
    editor = "Fang, Li-Zhi and Ruffini, R.",
    title = "{Cosmology for Grand Unified Theories with Radiatively Induced Symmetry Breaking}",
    reportNumber = "UPR-0185T",
    doi = "10.1103/PhysRevLett.48.1220",
    journal = "Phys. Rev. Lett.",
    volume = "48",
    pages = "1220--1223",
    year = "1982"
}

@article{Ivanov:1994pa,
    author = "Ivanov, P. and Naselsky, P. and Novikov, I.",
    title = "{Inflation and primordial black holes as dark matter}",
    reportNumber = "NORDITA-94-12-A",
    doi = "10.1103/PhysRevD.50.7173",
    journal = "Phys. Rev. D",
    volume = "50",
    pages = "7173--7178",
    year = "1994"
}

@article{Inoue:2001zt,
    author = "Inoue, Shogo and Yokoyama, Jun'ichi",
    title = "{Curvature perturbation at the local extremum of the inflaton's potential}",
    eprint = "hep-ph/0104083",
    archivePrefix = "arXiv",
    reportNumber = "OU-TAP-160",
    doi = "10.1016/S0370-2693(01)01369-7",
    journal = "Phys. Lett. B",
    volume = "524",
    pages = "15--20",
    year = "2002"
}

@article{Kinney:2005vj,
    author = "Kinney, William H.",
    title = "{Horizon crossing and inflation with large eta}",
    eprint = "gr-qc/0503017",
    archivePrefix = "arXiv",
    doi = "10.1103/PhysRevD.72.023515",
    journal = "Phys. Rev. D",
    volume = "72",
    pages = "023515",
    year = "2005"
}

@article{Kallosh:2019eeu,
    author = "Kallosh, Renata and Linde, Andrei",
    title = "{B-mode Targets}",
    eprint = "1906.04729",
    archivePrefix = "arXiv",
    primaryClass = "astro-ph.CO",
    doi = "10.1016/j.physletb.2019.134970",
    journal = "Phys. Lett. B",
    volume = "798",
    pages = "134970",
    year = "2019"
}

@article{Hawking:1971ei,
    author = "Hawking, Stephen",
    title = "{Gravitationally collapsed objects of very low mass}",
    doi = "10.1093/mnras/152.1.75",
    journal = "Mon. Not. Roy. Astron. Soc.",
    volume = "152",
    pages = "75",
    year = "1971"
}

@article{Carr:1974nx,
    author = "Carr, Bernard J. and Hawking, S. W.",
    title = "{Black holes in the early Universe}",
    doi = "10.1093/mnras/168.2.399",
    journal = "Mon. Not. Roy. Astron. Soc.",
    volume = "168",
    pages = "399--415",
    year = "1974"
}

@article{Sasaki:1986hm,
    author = "Sasaki, Misao",
    title = "{Large Scale Quantum Fluctuations in the Inflationary Universe}",
    reportNumber = "RRK-86-29",
    doi = "10.1143/PTP.76.1036",
    journal = "Prog. Theor. Phys.",
    volume = "76",
    pages = "1036",
    year = "1986"
}

@article{Mukhanov:1988jd,
    author = "Mukhanov, Viatcheslav F.",
    title = "{Quantum Theory of Gauge Invariant Cosmological Perturbations}",
    journal = "Sov. Phys. JETP",
    volume = "67",
    pages = "1297--1302",
    year = "1988"
}

@article{Planck:2018vyg,
    author = "Aghanim, N. and others",
    collaboration = "Planck",
    title = "{Planck 2018 results. VI. Cosmological parameters}",
    eprint = "1807.06209",
    archivePrefix = "arXiv",
    primaryClass = "astro-ph.CO",
    doi = "10.1051/0004-6361/201833910",
    journal = "Astron. Astrophys.",
    volume = "641",
    pages = "A6",
    year = "2020",
    note = "[Erratum: Astron.Astrophys. 652, C4 (2021)]"
}

@article{Planck:2018jri,
    author = "Akrami, Y. and others",
    collaboration = "Planck",
    title = "{Planck 2018 results. X. Constraints on inflation}",
    eprint = "1807.06211",
    archivePrefix = "arXiv",
    primaryClass = "astro-ph.CO",
    doi = "10.1051/0004-6361/201833887",
    journal = "Astron. Astrophys.",
    volume = "641",
    pages = "A10",
    year = "2020"
}

@article{Domenech:2021ztg,
    author = "Dom{\`e}nech, Guillem",
    title = "{Scalar Induced Gravitational Waves Review}",
    eprint = "2109.01398",
    archivePrefix = "arXiv",
    primaryClass = "gr-qc",
    doi = "10.3390/universe7110398",
    journal = "Universe",
    volume = "7",
    number = "11",
    pages = "398",
    year = "2021"
}

@article{Morse:2018kda,
    author = "Morse, Michael J. P. and Kinney, William H.",
    title = "{Large-$\eta$ constant-roll inflation is never an attractor}",
    eprint = "1804.01927",
    archivePrefix = "arXiv",
    primaryClass = "astro-ph.CO",
    doi = "10.1103/PhysRevD.97.123519",
    journal = "Phys. Rev. D",
    volume = "97",
    number = "12",
    pages = "123519",
    year = "2018"
}

@article{Gow:2020bzo,
    author = "Gow, Andrew D. and Byrnes, Christian T. and Cole, Philippa S. and Young, Sam",
    title = "{The power spectrum on small scales: Robust constraints and comparing PBH methodologies}",
    eprint = "2008.03289",
    archivePrefix = "arXiv",
    primaryClass = "astro-ph.CO",
    doi = "10.1088/1475-7516/2021/02/002",
    journal = "JCAP",
    volume = "02",
    pages = "002",
    year = "2021"
}

@article{AtacamaCosmologyTelescope:2025blo,
    author = "Louis, Thibaut and others",
    collaboration = "Atacama Cosmology Telescope",
    title = "{The Atacama Cosmology Telescope: DR6 power spectra, likelihoods and {\ensuremath{\Lambda}}CDM parameters}",
    eprint = "2503.14452",
    archivePrefix = "arXiv",
    primaryClass = "astro-ph.CO",
    reportNumber = "FERMILAB-PUB-25-0071-PPD",
    doi = "10.1088/1475-7516/2025/11/062",
    journal = "JCAP",
    volume = "11",
    pages = "062",
    year = "2025"
}

@article{SPT-3G:2025bzu,
    author = "Camphuis, E. and others",
    collaboration = "SPT-3G",
    title = "{SPT-3G D1: CMB temperature and polarization power spectra and cosmology from 2019 and 2020 observations of the SPT-3G main field}",
    eprint = "2506.20707",
    archivePrefix = "arXiv",
    primaryClass = "astro-ph.CO",
    reportNumber = "FERMILAB-PUB-25-0144-PPD",
    doi = "10.1103/7wt3-9v2y",
    journal = "Phys. Rev. D",
    volume = "113",
    number = "8",
    pages = "083504",
    year = "2026"
}

@inproceedings{Motohashi:2025qgd,
    author = "Motohashi, Hayato",
    title = "{Constant-Roll Inflation}",
    eprint = "2504.16757",
    archivePrefix = "arXiv",
    primaryClass = "astro-ph.CO",
    month = "4",
    year = "2025"
}

@article{Tzirakis:2007bf,
    author = "Tzirakis, Konstantinos and Kinney, William H.",
    title = "{Inflation over the hill}",
    eprint = "astro-ph/0701432",
    archivePrefix = "arXiv",
    doi = "10.1103/PhysRevD.75.123510",
    journal = "Phys. Rev. D",
    volume = "75",
    pages = "123510",
    year = "2007"
}

@article{Weinberg:2005vy,
    author = "Weinberg, Steven",
    title = "{Quantum contributions to cosmological correlations}",
    eprint = "hep-th/0506236",
    archivePrefix = "arXiv",
    reportNumber = "UTTG-01-05",
    doi = "10.1103/PhysRevD.72.043514",
    journal = "Phys. Rev. D",
    volume = "72",
    pages = "043514",
    year = "2005"
}

@article{Raveendran:2025pnz,
    author = "Raveendran, Rathul Nath",
    title = "{Validity of separate-universe approach in transient ultraslow-roll inflation}",
    eprint = "2506.23571",
    archivePrefix = "arXiv",
    primaryClass = "astro-ph.CO",
    doi = "10.1103/2gd5-xv73",
    journal = "Phys. Rev. D",
    volume = "112",
    number = "10",
    pages = "103507",
    year = "2025"
}

@ARTICLE{1899Natur..59..606G,
       author = {{Gibbs}, J. Willard},
        title = "{Fourier's Series}",
      journal = {Nature},
         year = 1899,
        month = apr,
       volume = {59},
       number = {1539},
        pages = {606},
          doi = {10.1038/059606a0},
       adsurl = {https://ui.adsabs.harvard.edu/abs/1899Natur..59..606G}
}

@article{Briaud:2025hra,
    author = {Briaud, Vadim and Karam, Alexandros and Koivunen, Niko and Tomberg, Eemeli and Veerm{\"a}e, Hardi and Vennin, Vincent},
    title = "{How deep is the dip and how tall are the wiggles in inflationary power spectra?}",
    eprint = "2501.14681",
    archivePrefix = "arXiv",
    primaryClass = "astro-ph.CO",
    doi = "10.1088/1475-7516/2025/05/097",
    journal = "JCAP",
    volume = "05",
    pages = "097",
    year = "2025"
}

@article{Vennin:2014xta,
    author = "Vennin, Vincent",
    title = "{Horizon-Flow off-track for Inflation}",
    eprint = "1401.2926",
    archivePrefix = "arXiv",
    primaryClass = "astro-ph.CO",
    doi = "10.1103/PhysRevD.89.083526",
    journal = "Phys. Rev. D",
    volume = "89",
    number = "8",
    pages = "083526",
    year = "2014"
}

@article{Hamada:2014iga,
    author = "Hamada, Yuta and Kawai, Hikaru and Oda, Kin-ya and Park, Seong Chan",
    title = "{Higgs Inflation is Still Alive after the Results from BICEP2}",
    eprint = "1403.5043",
    archivePrefix = "arXiv",
    primaryClass = "hep-ph",
    doi = "10.1103/PhysRevLett.112.241301",
    journal = "Phys. Rev. Lett.",
    volume = "112",
    number = "24",
    pages = "241301",
    year = "2014"
}

@article{Karam:2022nym,
    author = {Karam, Alexandros and Koivunen, Niko and Tomberg, Eemeli and Vaskonen, Ville and Veerm{\"a}e, Hardi},
    title = "{Anatomy of single-field inflationary models for primordial black holes}",
    eprint = "2205.13540",
    archivePrefix = "arXiv",
    primaryClass = "astro-ph.CO",
    doi = "10.1088/1475-7516/2023/03/013",
    journal = "JCAP",
    volume = "03",
    pages = "013",
    year = "2023"
}

@article{Cruces:2026qvl,
    author = "Cruces, Diego and He, Minxi and Pi, Shi and Wang, Jianing and Yamaguchi, Masahide and Zhu, Yuhang",
    title = "{Natura Non Facit Saltum: An Analytical Model of Smooth Slow-Roll to Ultra-Slow-Roll Transition}",
    eprint = "2603.17465",
    archivePrefix = "arXiv",
    primaryClass = "astro-ph.CO",
    month = "3",
    year = "2026"
}

@article{Ng:2021hll,
    author = "Ng, Kin-Wang and Wu, Yi-Peng",
    title = "{Constant-rate inflation: primordial black holes from conformal weight transitions}",
    eprint = "2102.05620",
    archivePrefix = "arXiv",
    primaryClass = "astro-ph.CO",
    doi = "10.1007/JHEP11(2021)076",
    journal = "JHEP",
    volume = "11",
    pages = "076",
    year = "2021"
}
\bibliographystyle{JHEP}

\end{document}